\documentclass[notitlepage]{article}
\usepackage[T1]{fontenc}
\usepackage[american]{babel}
\usepackage{amssymb}
\usepackage[dvips]{graphicx}
\usepackage{color}
\usepackage{vmargin}

\begin{document}

\title{AN INVARIANT\ FORMULATION\ OF\ SPECIAL\ RELATIVITY,\ OR\ THE\ ''TRUE\
TRANSFORMATION\ RELATIVITY,''\ AND\ ITS\ COMPARISON WITH EXPERIMENTS }
\author{Tomislav Ivezi\'{c} \\
\textit{Ru{%
\mbox
 {\it{d}\hspace{-.15em}\rule[1.25ex]{.2em}{.04ex}\hspace{-.05em}}}er Bo\v
{s}kovi\'{c} Institute, P.O.B. 180, 10002 Zagreb, Croatia}\\
\textit{\ ivezic@rudjer.irb.hr}}
\maketitle

\noindent Different formulations of special relativity are theoretically
discussed. First an invariant formulation, i.e., the ''true transformations
(TT) relativity,'' is exposed. There a physical quantity is represented by a
true tensor which comprises both components and a basis. Also the usual
covariant formulation and the ''apparent transformations (AT)\ relativity''
are considered. It is shown that all the experiments are in agreement with
the ''TT relativity'' but not always with the ''AT relativity.'' \medskip

\noindent PACS number(s): 03.30.+p


\section{INTRODUCTION}

In the recent papers \cite{tom1} and \cite{tom2} an invariant formulation of
special relativity (SR) is proposed and it is called the ''true
transformations (TT)\ relativity.'' Furthermore the differences between this
formulation, the usual covariant approach to SR and the traditionally used
''apparent transformations (AT) relativity'' (a typical example of the ''AT
relativity'' is Einstein's \cite{Einst} formulation of SR) are also examined
in \cite{tom1} and \cite{tom2}. Some parts of these formulations are
discussed in \cite{ivezic}, \cite{ive2} as well. The notions of the TT and
the AT are first introduced by Rohrlich \cite{rohrl1}, and, in the same
meaning, but not under that name, discussed in \cite{gamba} too. In \cite
{tom1,tom2} (and \cite{ivezic,ive2}) we have also presented the theoretical
discussion of the TT of the spacetime length for a moving rod and a moving
clock, and of the AT for the same examples, i.e., the AT of the spatial
distance, the Lorentz ''contraction,'' and the AT of the temporal distance,
the time ''dilatation.'' In this paper we expose the main theoretical
results from \cite{tom1,tom2,ivezic,ive2} and compare them with some
experimental results.

It is usually interpreted that the experiments on ''length contraction'' and
''time dilatation'' test SR, but the theoretical discussion from \cite
{tom1,tom2} shows that such an interpretation of the experiments refers
exclusively to the ''AT relativity,'' and not to the ''TT relativity.''

It has to be noted that in the experiments in the ''TT relativity,'' in the
same way as in the theory, see \cite{tom1,tom2}, the measurements in
different inertial frames of reference (IFRs) (and different
coordinatizations) have to refer to the same four-dimensional (4D) tensor
quantity. In the chosen IFR and the chosen coordinatization the measurement
of some 4D quantity has to contain the measurements of all parts of such a
quantity. However in almost all experiments that refer to SR only the
quantities belonging to the ''AT relativity'' were measured. From the ''TT
relativity'' viewpoint such measurements are incomplete, since only some
parts of a 4D quantity, not all, are measured. This fact presents a serious
difficulty in the reliable comparison of the existing experiments with the
''TT relativity,'' and, actually, we shall be able to compare in a
quantitative manner only some of the existing experiments with the ''TT
relativity.''

To examine the differences between the nonrelativistic theory, the commonly
used ''AT relativity,'' and the ''TT relativity'' we shall make the
comparison of these theories with some experiments in the following sections.

First in Sec. 2 we briefly expose the main theoretical results from \cite
{tom1,tom2} about the ''TT\ relativity'' and its theoretical comparison with
the ''AT relativity'' and with the usual covariant approach. In Sec. 4 we
discuss the ''muon'' experiment in the nonrelativistic approach, in the ''AT
relativity'' and in the ''TT relativity.'' Since the Michelson-Morley
experiment is discussed in detail in \cite{tom2} we expose in Secs. 5 and
5.1 only the main results from \cite{tom2} in order to use them for the
consideration of the modern laser versions in Sec. 5.2 and for the
discussion of the Kennedy-Thorndike type experiments in Sec. 6. In Secs. 7,
7.1 and 7.2 we consider different Iwes-Stillwel type experiments both in the
''AT relativity,'' Sec. 7.1, and in the ''TT relativity,'' Sec. 7.2. Finally
in Sec. 8 the discussion and conclusions are presented.

\section{A BRIEF THEORETICAL\ DISCUSSION\ OF\ THE\ THREE\ APPROACHES\ TO\ SR}

Rohrlich \cite{rohrl1}, and also Gamba \cite{gamba}, emphasized the role of\
the concept of sameness of a physical quantity for different observers. The
principal difference between the ''TT relativity'' and the ''AT relativity''
stems from the difference in that concept of sameness of a physical system,
i.e., of a physical quantity, for different observers. This concept of \emph{%
sameness} of a physical quantity for different observers actually determines
the difference in what is to be understood as a relativistic theory. Our
invariant approach to SR, i.e., the ''TT relativity,'' and the concept of
sameness of a physical quantity for different observers in that approach,
differs not only from the ''AT relativity'' approach but also from the usual
covariant approach (including \cite{rohrl1} and \cite{gamba}).

In the ''TT relativity'' SR is understood as the theory of a 4D spacetime
with pseudo-Euclidean geometry. All physical quantities (in the case when no
basis has been introduced) are described by true tensor fields, that are
defined on the 4D spacetime, and that satisfy true tensor equations
representing physical laws. When the coordinate system has been introduced
the physical quantities are mathematically represented by the
coordinate-based geometric quantities (CBGQs) that satisfy the
coordinate-based geometric equations. The CBGQs contain both the components
and the basis one-forms and vectors of the chosen IFR. Speaking in
mathematical language a tensor of type (k,l) is defined as a linear function
of k one-forms and l vectors (in old names, k covariant vectors and l
contravariant vectors) into the real numbers, see, e.g.,
\cite{Wald,Schutz,Misner}. If a coordinate system is chosen in some IFR then, in
general, any tensor quantity can be reconstructed from its components and
from the basis vectors and basis 1-forms of that frame, i.e., it can be
written in a coordinate-based geometric language, see, e.g., \cite{Misner}.
The symmetry transformations for the metric $g_{ab}$, i.e., the isometries 
\cite{Wald}, do not change $g_{ab}$; if we denote an isometry as $\Phi ^{*}$
then $(\Phi ^{*}g)_{ab}=g_{ab}.$ Thus an isometry leaves the
pseudo-Euclidean geometry of 4D spacetime of SR unchanged. At the same time
they do not change the true tensor quantities, or equivalently the CBGQs, in
physical equations. Thus \emph{isometries are }what Rohrlich \cite{rohrl1}
calls \emph{the TT}. In the ''TT relativity'' different coordinatizations of
an IFR are allowed and they are all equivalent in the description of
physical phenomena. Particularly two very different coordinatizations, the
Einstein (''e'') \cite{Einst} and ''radio'' (''r'') \cite{Leub}
coordinatization are discussed in \cite{tom1,tom2} and \cite{ive2} and will
be exploited in this paper as well. (In the ''e'' coordinatization the
Einstein synchronization \cite{Einst} of distant clocks and cartesian space
coordinates $x^{i}$ are used in the chosen IFR. The main features of the
''r'' coordinatization will be given below. For the recent discussion of the
conventionality of synchronization see \cite{ander} and references therein.)
The \emph{CBGQs }representing some 4D physical quantity in different
relatively moving IFRs, or in different coordinatizations of the chosen IFR, 
\emph{are all mathematically equal} since they are connected by the TT
(i.e., the isometries). Thus they are really the same quantity for different
observers, or in different coordinatizations. Hence \emph{in the ''TT
relativity'' the same quantity for different observers is either the true
tensor quantity or the CBGQ.} Therefore it is appropriate to call the ''TT
relativity'' approach (which deals with the true tensors or with the CBGQs)
as an invariant approach in contrast to the usual covariant approach (which
deals with the components of tensors taken in the ''e'' coordinatization).
We suppose that in the ''TT relativity'' such \emph{4D tensor quantities are
well-defined not only mathematically but also experimentally, as measurable
quantities with real physical meaning. The complete and well-defined
measurement from the ''TT relativity'' viewpoint is such measurement in
which all parts of some 4D quantity are measured.}

\emph{In the usual covariant approach} one does not deal with the true
tensors, or equivalently with CBGQs, but with \emph{the basis components of
tensors} (mainly in the ''e'' coordinatization) and with the equations of
physics written out in the component form. Mathematically speaking the
concept of a tensor in the usual covariant approach is defined entirely in
terms of the transformation properties of its components relative to some
coordinate system. Hence \emph{in the usual covariant approach the same
quantity for different observers is the component form of a true tensor, or
equivalently of a CBGQ, in some specific coordinatization.} The definitions
of the same quantity in \cite{rohrl1} and \cite{gamba} also refer to such
component form in the ''e'' coordinatization of tensor quantities and tensor
equations. Although it is true that the components of some tensor refer to
the same tensor quantity considered in two relatively moving IFRs $S$ and $%
S^{\prime }$ and in the ''e'' coordinatization, but they are not the same 4D
quantity since the bases are not included. This will be explicitly shown
below.

The third approach to SR uses the AT of some quantities. In contrast to the
TT (i.e., the isometries) the AT are not the transformations of spacetime
tensors and they do not refer to the same 4D quantity. \emph{The AT refer
exclusively to the component form of tensor quantities and in that form they
transform only some components of the whole tensor quantity.} In fact,
depending on the used AT, only a part of a 4D tensor quantity is transformed
by the AT. Such a part of a 4D quantity, when considered in different IFRs
(or in different coordinatizations of some IFR) corresponds to different
quantities in 4D spacetime. Some examples of\ the AT are: the AT of the
synchronously defined spatial length \cite{Einst}, i.e., the Lorentz
contraction, and the AT of the temporal distance, i.e., the conventional
dilatation of time that is introduced in \cite{Einst} and considered in \cite
{tom1,tom2}. Any formulation of SR which uses the AT we call the ''AT
relativity.'' An example of such formulation is Einstein's formulation of SR
which is based on his two postulates and which deals with all the mentioned
AT. Thus \emph{in the ''AT relativity'' the same quantity for different
observers is considered to be a part of a 4D tensor quantity which is
transformed by the AT.}

In this paper I use the same convention with regard to indices as in \cite
{tom1,tom2}. Repeated indices imply summation. Latin indices $a,b,c,d,...$
are to be read according to the abstract index notation, see \cite{Wald},
Sec.2.4.; they ''...should be viewed as reminders of the number and type of
variables the tensor acts on, \emph{not} as basis components.'' They
designate geometric objects in 4D spacetime. Thus, e.g., $l_{AB}^{a}$ (a
distance 4-vector $l_{AB}^{a}=x_{B}^{a}-x_{A}^{a}$ between two events $A$
and $B$ with the position 4-vectors $x_{A}^{a}$ and $x_{B}^{a}$) and $%
x_{A,B}^{a}$ are (1,0) tensors and they are defined independently of any
coordinate system. Greek indices run from 0 to 3, while latin indices $%
i,j,k,l,...$ run from 1 to 3, and they both designate the components of some
geometric object in some coordinate system, e.g., $x^{\mu }(x^{0},x^{i})$
and $x^{\mu ^{\prime }}(x^{0^{\prime }},x^{i^{\prime }})$ are two coordinate
representations of the position 4-vector $x^{a}$ in two different inertial
coordinate systems $S$ and $S^{\prime }.$ Similarly the metric tensor $%
g_{ab} $ denotes a tensor of type (0,2) (whose Riemann curvature tensor $%
R_{bcd}^{a} $ is everywhere vanishing; the spacetime of special relativity
is a flat spacetime, and this definition includes not only the IFRs but also
the accelerated frames of reference). This geometric object $g_{ab}$ is
represented in the component form in an IFR $S,$ and in the ''e''
coordinatization, i.e., in the $\left\{ e_{\mu }\right\} $ basis, by the $%
4\times 4$ diagonal matrix of components of $g_{ab}$, $g_{\mu \nu
,e}=diag(-1,1,1,1),$ and this is usually called the Minkowski metric tensor.
Note that the subscript $^{\prime }e^{\prime }$ stands for the Einstein
coordinatization.

In the following we shall also need the expression for the covariant 4D
Lorentz transformations $L^{a}{}_{b}$, which is independent of the chosen
synchronization, i.e., coordinatization of reference frames (see \cite{Fahn}%
, \cite{tom1,tom2} and \cite{ive2}). It is 
\begin{equation}
L^{a}{}_{b}\equiv L^{a}{}_{b}(v)=g^{a}{}_{b}-\frac{2u^{a}v_{b}}{c^{2}}+\frac{%
(u^{a}+v^{a})(u_{b}+v_{b})}{c^{2}(1+\gamma )},  \label{fah}
\end{equation}
where $u^{a}$ is the proper velocity 4-vector of a frame $S$ with respect to
itself, $u^{a}=cn^{a},$ $n^{a}$ is the unit 4-vector along the $x^{0}$ axis
of the frame $S,$ and $v^{a}$ is the proper velocity 4-vector of $S^{\prime
} $ relative to $S.$ Further $u\cdot v=u^{a}v_{a}$ and $\gamma =-u\cdot
v/c^{2}.$ When we use the Einstein coordinatization then $L^{a}{}_{b}$ is
represented by $L^{\mu }{}_{\nu ,e},$ the usual expression for pure Lorentz
transformation which connects two coordinate representations, basis
components (in the ''e'' coordinatization), $x_{e}^{\mu },$ $x_{e}^{\mu
^{\prime }}$ of a given event. $x_{e}^{\mu },$ $x_{e}^{\mu ^{\prime }}$
refer to two relatively moving IFRs (with the Minkowski metric tensors) $S$
and $S^{\prime },$ 
\begin{eqnarray}
x_{e}^{\mu ^{\prime }} &=&L^{\mu ^{\prime }}{}_{\nu ,e}x_{e}^{\nu },\qquad
\,L^{0^{\prime }}{}_{0,e}=\gamma _{e},\quad L^{0^{\prime
}}{}_{i,e}=L^{i^{\prime }}{}_{0,e}=-\gamma _{e}v_{e}^{i}/c,  \nonumber \\
L^{i^{\prime }}{}_{j,e} &=&\delta _{j}^{i}+(\gamma
_{e}-1)v_{e}^{i}v_{je}/v_{e}^{2},  \label{lorus}
\end{eqnarray}
where $v_{e}^{\mu }\equiv dx_{e}^{\mu }/d\tau =(\gamma _{e}c,\gamma
_{e}v_{e}^{i}),$ $d\tau \equiv dt_{e}/\gamma _{e}$ and $\gamma _{e}\equiv
(1-v_{e}^{2}/c^{2})^{1/2}$. Since $g_{\mu \nu ,e}$ is a diagonal matrix the
space $x_{e}^{i}$ and time $t_{e}$ $(x_{e}^{0}\equiv ct_{e})$ parts of $%
x_{e}^{\mu }$ do have their usual meaning.

The geometry of the spacetime is generally defined by the metric tensor $%
g_{ab},$ which can be expand in a coordinate basis in terms of its
components as $g_{ab}=g_{\mu \nu }dx^{\mu }\otimes dx^{\nu },$ and where $%
dx^{\mu }\otimes dx^{\nu }$ is an outer product of the basis 1-forms.

The connection between the basis vectors in the ''r'' and ''e''
coordinatizations is given as 
\begin{equation}
r_{0}=e_{0},\;r_{i}=e_{0}+e_{i},  \label{ere}
\end{equation}
see \cite{Leub}, \cite{ive2} and \cite{tom1,tom2}. The metric tensor $g_{ab}$
becomes $g_{ab}=g_{\mu \nu ,r}dx_{r}^{\mu }\otimes dx_{r}^{\nu }$ in the
coordinate-based geometric language and in the ''r'' coordinatization, where
the basis components of the metric tensor are 
\begin{equation}
g_{00,r}=g_{0i,r}=g_{i0,r}=g_{ij,r}(i\neq j)=-1,g_{ii,r}=0.  \label{geer}
\end{equation}
$dx_{r}^{\mu },$ $dx_{r}^{\nu }$ are the basis 1-forms in the ''r''
coordinatization and in $S,$ and $dx_{r}^{\mu }\otimes dx_{r}^{\nu }$ is an
outer product of the basis 1-forms, i.e., it is the basis for (0,2) tensors.

The transformation matrix $T^{\mu }{}_{\nu ,r}$ which transforms the tensor
quantities from the ''e'' coordinatization to the ''r'' coordinatization is
given as 
\begin{equation}
T^{\mu }{}_{\mu ,r}=-T^{0}{}_{i,r}=1,  \label{teer}
\end{equation}
and all other elements of $T^{\mu }{}_{\nu ,r}$ are $=0$. Using this $T^{\mu
}{}_{\nu ,r}$ we find 
\begin{equation}
x_{r}^{\mu }=T^{\mu }{}_{\nu ,r}x_{e}^{\nu },\quad
x_{r}^{0}=x_{e}^{0}-x_{e}^{1}-x_{e}^{2}-x_{e}^{3},\quad x_{r}^{i}=x_{e}^{i}.
\label{temini}
\end{equation}
For the sake of completeness we also quote the Lorentz transformation $%
L^{\mu ^{\prime }}\,_{\nu ,r}$ in the ''r'' coordinatization. It can be
easily found from $L^{a}{}_{b}$ (\ref{fah}) and the known $g_{\mu \nu ,r},$
and the elements that are different from zero are 
\begin{eqnarray}
x_{r}^{\prime \mu } &=&L^{\mu ^{\prime }}{}_{\nu ,r}x_{r}^{\nu },\quad
L^{0^{\prime }}{}_{0,r}=K,\quad L^{0^{\prime }}{}_{2,r}=L^{0^{\prime
}}{}_{3,r}=K-1,  \nonumber \\
L^{1^{\prime }}{}_{0,r} &=&L^{1^{\prime }}{}_{2,r}=L^{1^{\prime
}}{}_{3,r}=(-\beta _{r}/K),L^{1^{\prime }}{}_{1,r}=1/K,\quad L^{2^{\prime
}}{}_{2,r}=L^{3^{\prime }}{}_{3,r}=1,  \label{elr}
\end{eqnarray}
where $K=(1+2\beta _{r})^{1/2},$ and $\beta _{r}=dx_{r}^{1}/dx_{r}^{0}$ is
the velocity of the frame $S^{\prime }$ as measured by the frame $S$, $\beta
_{r}=\beta _{e}/(1-\beta _{e})$ and it ranges as $-1/2\prec \beta _{r}\prec
\infty .$

An example of isometry is the covariant 4D Lorentz transformation $%
L^{a}{}_{b}$ (\ref{fah}). When the coordinate basis is introduced then, for
example, the isometry $L^{a}{}_{b}$ (\ref{fah}) will be expressed as the
coordinate Lorentz transformation $L^{\mu ^{\prime }}{}_{\nu ,e}$ (\ref
{lorus}) in the ''e'' coordinatization, or as $L^{\mu ^{\prime }}\,_{\nu ,r}$
(\ref{elr}) in the ''r'' coordinatization.

Now we can better explain the above mentioned difference between three
approaches to SR in the understanding of the concept of the same quantity
for different observers. We shall consider some simple examples in the ''TT
relativity'': the spacetime length for a moving rod and then for a moving
clock. The same examples will be also examined in the ''AT relativity.''

\subsection{The spacetime length for a moving rod and a moving clock}

Let us take, for simplicity, to work in 2D spacetime. Then we consider a
true tensor quantity, a distance 4-vector (the (1,0) tensor) $%
l_{AB}^{a}=x_{B}^{a}-x_{A}^{a}$ between two events $A$ and $B$ (with the
position 4-vectors $x_{A}^{a}$ and $x_{B}^{a}$). $l_{AB}^{a}$ is chosen to
be a particular 4-vector which, in the usual ''3+1'' picture, corresponds to
an object, a rod, that is at rest in an IFR $S$ and situated along the
common $x_{e}^{1},x_{e}^{1^{\prime }}-$ axes. (The same example is already
considered in \cite{tom1,tom2} and \cite{ive2}.) This true tensor can be
represented in the coordinate-based geometric language in different bases, $%
\left\{ e_{\mu }\right\} $ and $\left\{ r_{\mu }\right\} $ in an IFR $S,$
and $\left\{ e_{\mu ^{\prime }}\right\} $ and $\left\{ r_{\mu ^{\prime
}}\right\} $ in a relatively moving IFR $S^{\prime },$ as $%
l_{AB}^{a}=l_{e}^{\mu }e_{\mu }=l_{r}^{\mu }r_{\mu }=l_{e}^{\mu ^{\prime
}}e_{\mu ^{\prime }}=l_{r}^{\mu ^{\prime }}r_{\mu ^{\prime }},$ where, e.g., 
$e_{\mu }$ are the basis 4-vectors, $e_{0}=(1,0,0,0)$ and so on, and $%
l_{e}^{\mu }$ are the basis components when the ''e'' coordinatization is
chosen in some IFR $S.$ The decompositions $l_{e}^{\mu }e_{\mu }$ and $%
l_{r}^{\mu }r_{\mu }$ (in an IFR $S,$ and in the ''e'' and ''r''
coordinatizations respectively) and $l_{e}^{\mu ^{\prime }}e_{\mu ^{\prime
}} $ and $l_{r}^{\mu ^{\prime }}r_{\mu ^{\prime }}$ (in a relatively moving
IFR $S^{\prime }$, and in the ''e'' and ''r'' coordinatizations
respectively) of the true tensor $l_{AB}^{a}$ are all mathematically \emph{%
equal} quantities. Thus they are really the same quantity considered in
different relatively moving IFRs and in different coordinatizations. (The
expressions for $l_{r}^{\mu }$ and $l_{r}^{\mu ^{\prime }}$ can be easily
found from the known transformation matrix $T_{\;\nu ,r}^{\mu }.$)
Particularly for this choice of the geometric quantity $l_{AB}^{a}$ its
decomposition in the ''e'' coordinatization and in $S$ is $%
l_{AB}^{a}=l_{e}^{0}e_{0}+l_{e}^{1}e_{1}=0e_{0}+L_{0}e_{1},$ while in $%
S^{\prime },$ where the rod is moving, it becomes $l_{AB}^{a}=-\beta
_{e}\gamma _{e}L_{0}e_{0^{\prime }}+\gamma _{e}L_{0}e_{1^{\prime }},$ and,
as explained above, it holds that 
\begin{equation}
l_{AB}^{a}=0e_{0}+L_{0}e_{1}=-\beta _{e}\gamma _{e}L_{0}e_{0^{\prime
}}+\gamma _{e}L_{0}e_{1^{\prime }}.  \label{trucon}
\end{equation}
We see from (\ref{trucon}) that in the ''e'' coordinatization there is a
dilatation of the spatial part $l_{e}^{1^{\prime }}=\gamma _{e}L_{0}$ with
respect to $l_{e}^{1}=L_{0}.$ Hovewer it is clear from the above discussion
that comparison of only spatial parts of the components of the distance
4-vector $l_{AB}^{a}$ in $S$ and $S^{\prime }$ is physically meaningless in
the ''TT relativity.'' When only some components of the whole tensor
quantity are taken alone then they do not represent some definite physical
quantity in the 4D spacetime. Similarly the decompositions of $l_{AB}^{a}$
in the ''r'' cordinatization are 
\begin{equation}
l_{AB}^{a}=-L_{0}r_{0}+L_{0}r_{1},=-KL_{0}r_{0^{\prime }}+(1+\beta
_{r})(1/K)L_{0}r_{1^{\prime }},  \label{trur}
\end{equation}
where $K=(1+2\beta _{r})^{1/2}.$ \emph{In the ''TT relativity'' the
geometric quantity }$l_{AB}^{a},$\emph{\ i.e., the coordinate-based
geometric quantities }$l_{e}^{\mu }e_{\mu }=l_{e}^{\mu ^{\prime }}e_{\mu
^{\prime }}=l_{r}^{\mu }r_{\mu }=l_{r}^{\mu ^{\prime }}r_{\mu ^{\prime }},$ 
\emph{comprising both, components and the basis, is the same 4D quantity for
different observers. }Note that if $l_{e}^{0}=0$ then $l_{e}^{\mu ^{\prime
}} $ in any other IFR $S^{\prime }$ will contain the time component $%
l_{e}^{0^{\prime }}\neq 0.$ The spacetime length $l$ between two points
(events) in 4D spacetime is defined as 
\begin{equation}
l=(g_{ab}l^{a}l^{b})^{1/2}.  \label{elspat}
\end{equation}
This spacetime length (\ref{elspat}) is frame and coordinatization
independent quantity, i.e., it holds that $l=(l_{e,r}^{\mu }l_{\mu
e,r})^{1/2}=(l_{e,r}^{\mu ^{\prime }}l_{\mu ^{\prime }e,r})^{1/2}=L_{0}.$ In
the ''e'' coordinatization the geometrical quantity $l^{2}$ can be written
in terms of its representation $l_{e}^{2},$ with the separated spatial and
temporal parts, $l^{2}=l_{e}^{2}=(l_{e}^{i}l_{ie})-(l_{e}^{0})^{2}$. Such
separation remains valid in other inertial coordinate systems with the
Minkowski metric tensor, and in $S^{\prime }$ one finds $l^{2}=l_{e}^{\prime
2}=(l_{e}^{i^{\prime }}l_{i^{\prime }e})-(l_{e}^{0^{\prime }})^{2},$ where $%
l_{e}^{\mu ^{\prime }}$ in $S^{\prime }$ is connected with $l_{e}^{\mu }$ in 
$S$ by the Lorentz transformation $L^{\mu ^{\prime }}{}_{\nu ,e}$ (\ref
{lorus}). Further in the ''e'' coordinatization and in $S,$ the rest frame
of the rod, where the temporal part of $l_{e}^{\mu }$ is $l_{e}^{0}=0,$ the
spacetime length $l$ is a measure of the spatial distance, i.e., of the rest
spatial length of the rod, as in the prerelativistic physics. Since $g_{\mu
\nu ,r},$ in contrast to $g_{\mu \nu ,e},$ is not a diagonal matrix, then in 
$l_{r}^{2}$ (the representation of $l^{2}$ in the ''r'' coordinatization)
the spatial and temporal parts are not separated.

In a similar manner we can choose another particular choice for the distance
4-vector $l_{AB}^{a},$ which will correspond to the well-known ''muon
experiment,'' and which is interpreted in the ''AT relativity'' in terms of
the time ''dilatation''. (This example is also investigated in \cite
{tom1,tom2}.) First we consider this example in the ''TT relativity.'' The
distance 4-vector $l_{AB}^{a}$ will be examined in two relatively moving
IFRs $S$ and $S^{\prime }$, i.e., in the $\left\{ e_{_{\mu }}\right\} $ and $%
\left\{ e_{\mu ^{\prime }}\right\} $ bases. The $S$ frame is chosen to be
the rest frame of the muon. Two events are considered; the event $A$
represents the creation of the muon and the event $B$ represents its decay
after the lifetime $\tau _{0}$ in $S.$ The position 4-vectors of the events $%
A$ and $B$ in $S$ are taken to be on the world line of a standard clock that
is at rest in the origin of $S.$ The distance 4-vector $%
l_{AB}^{a}=x_{B}^{a}-x_{A}^{a}$ that connects the events $A$ and $B$ is
directed along the $e_{0}$ basis vector from the event $A$ toward the event $%
B.$ This geometric quantity can be written in the coordinate-based geometric
language. Thus it can be decomposed in the bases $\left\{ e_{\mu }\right\} $
and $\left\{ e_{\mu ^{\prime }}\right\} $ as 
\begin{equation}
l_{AB}^{a}=c\tau _{0}e_{0}+0e_{1}=\gamma c\tau _{0}e_{0}^{\prime }-\beta
\gamma c\tau _{0}e_{1}^{\prime }.  \label{comu}
\end{equation}
and similarly in the ''r'' coordinatization as 
\begin{equation}
l_{AB}^{a}==c\tau _{0}r_{0}+0r_{1}=Kc\tau _{0}r_{0}^{\prime }-\beta
_{r}K^{-1}c\tau _{0}r_{1}^{\prime }.  \label{coer}
\end{equation}
We again see that these decompositions, containing both the basis components
and the basis vectors, are the same geometric quantity $l_{AB}^{a}.$ $%
l_{AB}^{a}$ does have only temporal parts in $S$, while in the $\left\{
e_{\mu ^{\prime }}\right\} $ basis $l_{AB}^{a}$ contains not only the
temporal part but also the spatial part. The spacetime length $l$ is always
a well-defined quantity in the ''TT relativity'' and for this example it is $%
l=(l_{e}^{\mu }l_{\mu e})^{1/2}=(l_{e}^{\mu ^{\prime }}l_{\mu ^{\prime
}e})^{1/2}=(l_{r}^{\mu }l_{\mu r})^{1/2}=(l_{r}^{\mu ^{\prime }}l_{\mu
^{\prime }r})^{1/2}=(-c^{2}\tau _{0}^{2})^{1/2}$. Since in $S$ the spatial
parts $l_{e,r}^{1}$ of $l_{e,r}^{\mu }$ are zero the spacetime length $l$ in 
$S$ is a measure of the temporal distance, as in the prerelativistic
physics; one defines that $c^{2}\tau _{0}^{2}=-l_{e}^{\mu }l_{\mu
e}=-l_{r}^{\mu }l_{\mu r}.$

These examples provide a nice possibility to discover the difference in the
concept of the same quantity for different observers between the ''TT
relativity'' and the usual covariant approach to SR. The usual covariant
approach does not consider the true tensor quantity, e.g., the distance
4-vector $l_{AB}^{a}$ (or equivalently the CBGQ $l_{e}^{\mu }e_{\mu },$
etc.), but only the basis components, $l_{e}^{\mu }$ and $l_{e}^{\nu
^{\prime }},$ in the ''e'' coordinatization. The \emph{basis components
(e.g., }$l_{e}^{\mu }$\emph{\ and }$l_{e}^{\nu ^{\prime }}$\emph{) are
considered to be the same quantity for different observers from the point of
view of the usual covariant approach to SR.} However, in contrast to the
above equalities for the CBGQs, the sets of components, $l_{e}^{\mu }$ and $%
l_{e}^{\nu ^{\prime }},$ taken alone, are not equal, $l_{e}^{\mu }\neq
l_{e}^{\nu ^{\prime }},$ and thus they are not the same quantity from the
''TT relativity'' viewpoint. From the mathematical point of view the
components of, e.g., a $(1,0)$ tensor are its values (real numbers) when the
basis one-form, for example, $e^{\alpha },$ is its argument (see, e.g., \cite
{Schutz}). Thus, for example, $l_{AB}^{a}(e^{\alpha })=l_{e}^{\mu }e_{\mu
}(e^{\alpha })=l_{e}^{\alpha }$ (where $e^{\alpha }$ is the basis one-form
in an IFR $S$ and in the ''e'' coordinatization), while $l_{AB}^{a}(e^{%
\alpha ^{\prime }})=l_{e}^{\mu ^{\prime }}e_{\mu ^{\prime }}(e^{\alpha
^{\prime }})=l_{e}^{\alpha ^{\prime }}$ (where $e^{\alpha ^{\prime }}$ is
the basis one-form in $S^{\prime }$ and in the ''e'' coordinatization).
Obviously $l_{e}^{\alpha }$ and $l_{e}^{\alpha ^{\prime }}$ are not the same
real numbers since the basis one-forms $e^{\alpha }$ and $e^{\alpha ^{\prime
}}$ are different bases. It is true that the components of some tensor \emph{%
refer }to the same tensor quantity considered in two relatively moving IFRs $%
S$ and $S^{\prime }$ and in the ''e'' coordinatization, but \emph{they are
not equal }since the bases are not included.

\subsection{The\ ''AT\ relativity''\ and\ the AT\ of special and temporal
distances}

As already said the AT refer exclusively to the component form of tensor
quantities and in that form they transform only \emph{some components }of
the whole tensor quantity.\emph{\ }Such a part of a 4D quantity, when
considered in different IFRs (or in different coordinatizations of some
IFR), corresponds to different quantities in 4D spacetime. The usual, i.e.,
Einstein's formulation of SR is based on two postulates: the principle of
relativity and the postulate that the coordinate, one-way, speed of light is
isotropic and constant. In that formulation the AT of the synchronously
defined \emph{spatial length} \cite{Einst} and the AT of \emph{the temporal
distance} \cite{Einst} are considered as the main ''relativistic''
consequences of the postulates. Namely the Lorentz transformations are
derived from the two mentioned postulates and then the Lorentz contraction
and the dilatation of time are interpreted as that they are the Lorentz
transformations of spatial and temporal distances. However the Lorentz
transformations are the TT, as can be seen from the preceding sections; they
always transform the whole 4D tensor quantity and thus they refer to the
same quantity in 4D spacetime, see, e.g., the relations (\ref{trucon}) and (%
\ref{comu}), or (\ref{trur}) and (\ref{coer}). Since the Lorentz
transformations are the TT, i.e., the isometries, they also do not change
the pseudo-Euclidean geometry of the spacetime. On the other hand, as will
be shown below, the Lorentz contraction and the dilatation of time are
typical examples of the AT. The Einstein formulation of SR uses the AT,
e.g., the Lorentz contraction and the dilatation of time, as important
ingredients of the theory (and also in experimental testing of the theory).
Any formulation of SR, which uses some of the AT, we call the ''AT
relativity.''

In order to better explain the difference between the TT and the AT we now
consider the same two examples as above but from the point of view of the
conventional, i.e., Einstein's \cite{Einst} interpretations of \emph{the} 
\emph{spatial length} of the moving rod and \emph{the temporal distance} for
the moving clock. These examples are already considered in \cite{tom1,tom2}
and \cite{ive2} and here we only quote the main results and the definitions.

The synchronous definition of \emph{the spatial length}, introduced by
Einstein \cite{Einst}, defines length as \emph{the spatial distance} between
two spatial points on the (moving) object measured by simultaneity in the
rest frame of the observer. The concept of sameness of a physical quantity
is quite different in the ''AT relativity'' but in the ''TT relativity.''
Indeed, in the usual ''AT relativity'' one takes only some basis components
of the whole 4D tensor quantity $l_{AB}^{a}$\ (that is, of the CBGQs $%
l_{e}^{\mu }e_{\mu }$\ and $l_{e}^{\mu ^{\prime }}e_{\mu ^{\prime }}$) in $S$%
\ and $S^{\prime },$\ then performs some additional manipulations with them,
and considers that the constructed quantities represent the same physical
quantity for observers in two relatively moving IFRs $S$\ and $S^{\prime }$.
Thus for the Einstein's definition of \emph{the spatial length} one
considers only \emph{the component} $l_{e}^{1}=L_{0}$ of $l_{e}^{\mu }e_{\mu
}$ (when $l_{e}^{0}$ is taken $=0,$ i.e., the spatial ends of the rod at
rest in $S$ are taken simultaneously at $t=0$) and compares it with the
quantity which is obtained in the following way; first one performs the
Lorentz transformation $L^{\mu }{}_{\nu ^{\prime },e}$ of the basis
components $l_{e}^{\mu ^{\prime }}$ (but not of the basis itself) from $%
S^{\prime }$ to $S,$ which yields 
\begin{eqnarray}
l_{e}^{0} &=&\gamma _{e}l_{e}^{0^{\prime }}+\gamma _{e}\beta
_{e}l_{e}^{1^{\prime }}  \nonumber \\
l_{e}^{1} &=&\gamma _{e}l_{e}^{1^{\prime }}+\gamma _{e}\beta
_{e}l_{e}^{0^{\prime }}.  \label{elcon}
\end{eqnarray}
Then one retains only the transformation of the spatial component $l_{e}^{1}$
(the second equation in (\ref{elcon})) \emph{neglecting completely the
transformation of the temporal part} $l_{e}^{0}$ (the first equation in (\ref
{elcon})). Furthermore in the transformation for $l_{e}^{1}$ one takes that
the temporal part in $S^{\prime }$ $l_{e}^{0^{\prime }}=0,$ ( i.e., the
spatial ends of the rod moving in $S^{\prime }$ are taken simultaneously at
some \emph{arbitrary }$t^{\prime }=b$). The quantity obtained in such a way
will be denoted as $L_{e}^{1^{\prime }}$ (it is not equal to $%
l_{e}^{1^{\prime }}$ appearing in the transformation equations (\ref{elcon}%
)) This quantity $L_{e}^{1^{\prime }}$ defines in the ''AT relativity'' 
\emph{the synchronously determined spatial length }of the moving rod in $%
S^{\prime }$. The mentioned procedure gives $l_{e}^{1}=\gamma
_{e}L_{e}^{1^{\prime }},$ that is, the famous formula for the Lorentz
contraction, 
\begin{equation}
L_{e}^{1^{\prime }}=l_{e}^{1}/\gamma _{e}=L_{0}/\gamma _{e},  \label{apcon}
\end{equation}
This quantity, $L_{e}^{1^{\prime }}=L_{0}/\gamma _{e},$ is the usual Lorentz
contracted \emph{spatial length}$,$ and the quantities $L_{0}$ and $%
L_{e}^{1^{\prime }}$ are considered in the ''AT relativity'' to be the same
quantity for observers in $S$ and $S^{\prime }$. The comparison with the
relation (\ref{trucon}) clearly shows that constructed quantities $L_{0}$%
\emph{\ and }$L_{e}^{1^{\prime }}$\emph{\ are two different and independent
quantities in 4D spacetime.} Namely, these quantities are obtained by the
same measurements in $S$ and $S^{\prime };$ the spatial ends of the rod are
measured simultaneously at some $t_{e}=a$ in $S$ and also at some $%
t_{e}^{\prime }=b$ in $S^{\prime }$; $a$ in $S$ and $b$ in $S^{\prime }$ are
not related by the Lorentz transformation $L^{\mu }{}_{\nu ,e}$ or any other
coordinate transformation. Thus, \emph{in the ''TT relativity'' the same
quantity for different observers is the tensor quantity, the 4-vector }$%
l_{AB}^{a}=l_{e}^{\mu }e_{\mu }=l_{e}^{\mu ^{\prime }}e_{\mu ^{\prime
}}=l_{r}^{\mu }r_{\mu }=l_{r}^{\mu ^{\prime }}r_{\mu ^{\prime }};$\emph{\
only one quantity in 4D spacetime.} However \emph{in the ''AT relativity''
different quantities in 4D spacetime, the spatiall distances }$l_{e}^{1},$%
\emph{\ }$L_{e}^{1^{\prime }}$\emph{\ (or in the ''r'' coordinatization }$%
l_{r}^{1},$\emph{\ }$L_{r}^{1^{\prime }}$\emph{)\ are considered as the same
quantity for different observers.} The relation for the Lorentz
''contraction'' of the moving rod in the ''r'' coordinatization can be
easily obtained performing the same procedure as in the ''e''
coordinatization, and it is 
\begin{equation}
L_{r}^{1^{\prime }}=L_{0}/K=(1+2\beta _{r})^{-1/2}L_{0},  \label{aper}
\end{equation}
see also \cite{tom1,tom2} and \cite{ive2}. We see from (\ref{aper}) that
there is a length dilatation $\infty \succ L_{r}^{1^{\prime }}\succ L_{0}$
for $-1/2\prec \beta _{r}\prec 0$ and the standard length ''contraction'' $%
L_{0}\succ L_{r}^{1^{\prime }}\succ 0$ for positive $\beta _{r},$ which
clearly shows that the ''Lorentz contraction'' is not physically correctly
defined transformation. \emph{Thus the Lorentz contraction is the
transformation that connects different quantities (in 4D spacetime) in }$S$%
\emph{\ and }$S^{\prime },$\emph{\ or in different coordinatizations, which
implies that it is - an AT. }

The same example of the ''muon decay'' will be now considered in the ''AT
relativity'' (see also \cite{tom1,tom2}). In the ''e'' coordinatization the
events $A$ and $B$ are again on the world line of a muon that is at rest in $%
S.$ We shall see once again that the concept of sameness of a physical
quantity is quite different in the ''AT relativity.'' Thus for this example
one compares \emph{the basis component} $l_{e}^{0}=c\tau _{0}$ of $%
l_{e}^{\mu }e_{\mu }$ with the quantity, which is obtained from \emph{the
basis component} $l_{e}^{0^{\prime }}$ in the following manner; first one
performs the Lorentz transformation of the basis components $l_{e}^{\mu }$
(but not of the basis itself) from the muon rest frame $S$ to the frame $%
S^{\prime }$ in which the muon is moving. This procedure yields 
\begin{eqnarray}
l_{e}^{0^{\prime }} &=&\gamma _{e}l_{e}^{0}-\gamma _{e}\beta _{e}l_{e}^{1} 
\nonumber \\
l_{e}^{1^{\prime }} &=&\gamma _{e}l_{e}^{1}-\gamma _{e}\beta _{e}l_{e}^{0}.
\label{eltime}
\end{eqnarray}
Similarly as in the Lorentz contraction \emph{one now forgets the
transformation of the spatial part} $l_{e}^{1^{\prime }}$ (the second
equation in (\ref{eltime})) and considers only the transformation of the
temporal part $l_{e}^{0^{\prime }}$ (the first equation in (\ref{eltime})).
This is, of course, an incorrect step from the ''TT relativity'' viewpoint.
Then taking that $l_{e}^{1}=0$ (i.e., that $x_{Be}^{1}=x_{Ae}^{1}$) in the
equation for $l_{e}^{0^{\prime }}$ (the first equation in (\ref{eltime}))
one finds the new quantity which will be denoted as $L_{e}^{0^{\prime }}$
(it is not the same as $l_{e}^{0^{\prime }}$ appearing in the transformation
equations (\ref{eltime})). The temporal distance $l_{e}^{0}$ defines in the
''AT relativity,'' and in the ''e'' coordinatization, the muon lifetime at
rest, while $L_{e}^{0^{\prime }}$ is considered in the ''AT relativity,''
and in the ''e'' coordinatization, to define the lifetime of the moving muon
in $S^{\prime }.$ The relation connecting $L_{e}^{0^{\prime }}$ with $%
l_{e}^{0},$ which is obtained by the above procedure, is then the well-known
relation for the time ''dilatation,'' 
\begin{equation}
L_{e}^{0^{\prime }}/c=t_{e}^{\prime }=\gamma _{e}l_{e}^{0}/c=\tau
_{0}(1-\beta _{e}^{2})^{-1/2}.  \label{tidil}
\end{equation}
By the same procedure we can find (see also \cite{tom1,tom2}) the relation
for the time ''dilatation'' in the ''r'' coordinatization 
\begin{equation}
L_{r}^{0^{\prime }}=Kl_{r}^{0}=(1+2\beta _{r})^{1/2}c\tau _{0}.
\label{tider}
\end{equation}
This relation shows that the new quantity $L_{r}^{0^{\prime }},$ which
defines in the ''AT relativity'' the temporal separation in $S^{\prime },$
where the clock is moving, is smaller - ''time contraction'' - than the
temporal separation $l_{r}^{0}=c\tau _{0}$ in $S,$ where the clock is at
rest, for $-1/2\prec \beta _{r}\prec 0,$ and it is larger - ''time
dilatation'' - for $0\prec \beta _{r}\prec \infty $. From this consideration
we conclude that \emph{in the ''TT relativity'' the same quantity for
different observers is the tensor quantity, the 4-vector }$%
l_{AB}^{a}=l_{e}^{\mu }e_{\mu }=l_{e}^{\mu ^{\prime }}e_{\mu ^{\prime
}}=l_{r}^{\mu }r_{\mu }=l_{r}^{\mu ^{\prime }}r_{\mu ^{\prime }};$\emph{\
only one quantity in 4D spacetime.} However \emph{in the ''AT relativity''
different quantities in 4D spacetime, the temporal distances }$l_{e}^{0},$%
\emph{\ }$L_{e}^{0^{\prime }},$\emph{\ }$l_{r}^{0},$\emph{\ }$%
L_{r}^{0^{\prime }}$ \emph{are considered as the same quantity for different
observers. This shows that the time ''dilatation'' is the transformation
connecting different quantities (in 4D spacetime) in }$S$\emph{\ and }$%
S^{\prime },$\emph{\ or in different coordinatizations, which implies that
it is - an AT. }

The consideration performed in the preceding sections and in this section
reveals that the basic elements of the ''TT relativity,'' as an
''invariant'' formulation of SR, and of the usual Einstein formulation of
SR, as an ''AT relativity'' formulation, are quite different. Einstein's
formulation is based on two postulates: (i) the principle of relativity and
(ii) the postulate that the coordinate, one-way, speed of light is isotropic
and constant. In the ''TT relativity'' the primary importance is attributed
to the geometry of the spacetime; it is supposed that the geometry of our 4D
spacetime is a pseudo-Euclidean geometry. The physical quantities are
represented by geometric quantities, either by true tensors (when no basis
is chosen) or equivalently (when the coordinate basis is introduced) by the
CBGQs. Thence in the ''TT relativity'' there is no need to postulate the
principle of relativity as a fundamental law. It is replaced by the
requirement that the physical laws must be expressed as true tensor
equations or equivalently as the coordinate-based geometric equations in the
4D spacetime. Since the ''TT relativity'' deals on the same footing with all
possible coordinatizations of a chosen reference frame then the second
Einstein postulate (ii) also does not hold, in general, in the ''TT
relativity.'' Namely, as we have remarked earlier, only in Einstein's
coordinatization the coordinate, one-way, speed of light is isotropic and
constant, while in, e.g., the ''r'' coordinatization, it is not the case.

In numerous textbooks and papers the Lorentz contraction and the dilatation
of time are considered as very important ''relativistic effects.'' In the
discussions about these effects it is always understood that the coordinate
Lorentz transformation $L^{\mu ^{\prime }}{}_{\nu ,e}$ (\ref{lorus}) in the
''e'' coordinatization transforms the rest length $L_{0}$ to the Lorentz
contracted length $L_{e}^{1^{\prime }}$, i.e., the formula for the Lorentz
contraction (\ref{apcon}) is interpreted as the Lorentz transformation of
the synchronously determined rest length $L_{0}.$ Similarly happens with the
formula for the time dilatation (\ref{tidil}), which is interpreted as the
Lorentz transformation of the proper time interval $\tau _{0}$ (both events
happen at the same spatial point) to the time interval $L_{e}^{0^{\prime
}}/c $ in the moving frame in which these events happen at different spatial
points. Our consideration about the spacetime length and the AT of spatial
and temporal distances reveals that \emph{the Lorentz contraction and the
dilatation of time are the AT and have nothing to do with the Lorentz
transformation as the TT. }Thus \emph{the Lorentz contraction and the
dilatation of time }are certainly not true relativistic transformations, or
to be more precise, they \emph{have nothing in common with SR. }They surely
are not important relativistic effects. Already in 1967. Gamba \cite{gamba}
clearly stated for the Lorentz contraction: ''Although it is completely
useless concept in physics, it will probably continue to remain in the books
as an historical relic for the fascination of the layman.'' From our
consideration follows that the same can be said for the dilatation of time.
However, what is really surprising, after more than thirty years from
Rohrlich's paper \cite{rohrl1} and Gamba's paper \cite{gamba} the Lorentz
contraction and the dilatation of time are still intensively investigated
theoretically and experimentally as \emph{relativistic effects} in numerous
scientific papers and books. It is generally believed that the apparatus for
high-energy experiments in particle physics are aready designed in such a
way that they take into account longer decay time (the dilatation of time)
for moving particle. In the leading physical journals, e.g., in Physical
Review C under the heading - Relativistic Nuclear Collisions, one can
permanently encounter theoretical and experimental articles in which the
Lorentz contraction is understood as an essential part of the relativistic
theory. Thus, for example, it is generally accepted in ultra-relativistic
nuclear collisions, see, e.g., \cite{geiger}: ''that in the center-of-mass
frame two \emph{highly Lorentz contracted nuclei} (my emphasis) pass through
each other .... .'' Also it is taken in ultrarelativistic heavy-ion
reactions that, e.g., \cite{borch}: ''While the longitudinal extension of
the \emph{valence quarks} in a fast-moving nucleon \emph{does indeed look
Lorentz contracted }(my emphasis) to a stationary observer in the usual
way... .'' This issue of ultra-relativistic nuclear collisions will be
discussed in more detail elsewhere.

\subsection{The discussion of some other definitions of the spatial length}

Next we consider two other definitions of the \emph{spatial length}. The
first one is an ''asynchronous'' definition, see, e.g., \cite{caval} and 
\cite{gren} and the references therein. (Actually one can speak about the
asynchronous formulation of SR.) According to the asynchronous description
the spatial length of a moving body is defined as the spatial distance
between two points on it, as measured by simultaneity in the rest frame of
the body. Namely in the asynchronous formulation of SR the distance 4-vector 
$l_{AB}^{a}=x_{B}^{a}-x_{A}^{a}$ between two events $A$ and $B$ (with the
position 4-vectors $x_{A}^{a}$ and $x_{B}^{a}$) is written only in the
component form and in the ''e'' coordinatization. In $S,$ the rest frame of
the body, it is (in 2D spacetime) $l_{AB}^{\mu }=(0,L_{0})$ ($L_{0}$ is the
rest length and it is determined synchronously in $S$). In $S^{\prime },$
where the body is moving, the component form in the ''e'' coordinatization
of $l_{AB}^{a}$ is $l_{AB}^{\mu ^{\prime }}=(-\beta _{e}\gamma
_{e}L_{0},\gamma _{e}L_{0}).$ Now comes the main point in the asynchronous
definition. It is interpreted in the asynchronous formulation of SR that the
spatial part $l_{AB}^{1^{\prime }}=\gamma _{e}L_{0}=L^{\prime }$ of $%
l_{AB}^{\mu ^{\prime }}$ is the ''asynchronous'' \emph{length} $L^{\prime }$%
, determined asynchronously (since the temporal part is $\neq 0$), in the
frame $S^{\prime }$ in which the body is moving. One can say that there is a
Lorentz lengthening in the asynchronous formulation, instead of the usual
Lorentz contraction that exists in the ''synchronous,'' i.e., the Einstein
formulation of SR. It is considered in the asynchronous formulation that $%
L^{\prime }$ in $S^{\prime }$ and $L_{0}$ in $S$ refer to the same quantity.
The common feature for both formulations is that \emph{the spatial length of
a moving body is assumed to be a well defined physical quantity in 4D
spacetime.} Our formulation with true tensors (or the CBGQs) reveals that
this is not true; a well defined physical quantity in 4D spacetime that is
connected with a moving body can be only a 4D tensor quantity, e.g., either
the spacetime length $l$ (\ref{elspat}), or the distance 4-vector $%
l_{AB}^{a}=x_{B}^{a}-x_{A}^{a}.$ If, for example, one does not use the ''e''
coordinatization but the ''r'' coordinatization, then both formulations
(synchronous and asynchronous), which deal with the spatial length as a well
defined physical quantity, become meaningless. It is clear from the
discussion in Secs. 2 and 2.1 that comparison of only spatial (or temporal)
parts of the components of the distance 4-vector $l_{AB}^{a}$ in $S$ and $%
S^{\prime }$ is physically meaningless in the ''TT relativity,'' since \emph{%
some components of a 4D tensor quantity, when they are taken alone, do not
actually represent any 4D physical quantity.} Also we remark that the whole
tensor quantity $l_{AB}^{a}$\ comprising components and the basis is
transformed by the Lorentz transformation from $S$\ to $S^{\prime }.$ This
discussion shows that the asynchronous formulation of SR also belongs to the
''AT\ relativity.''

The next definition which will be examined is the relativistic (or radar)
length \cite{strel}. (One can speak about the radar formulation of SR.) It
is assumed in \cite{strel} that the relativistic length (the length of a
fast-moving rod) is defined as (the third article in \cite{strel}): ''the
half-sum of distances covered by a light signal in direct and opposite
directions along the rod.'' In the 4D spacetime Strel'tsov defines the
4-vector of relativistic length $l_{rel}^{\mu }$ (actually this length is
not the 4-vector but it is the component form in the ''e'' coordinatization
of a 4-vector) as: ''the half-difference of two light 4-vectors (i.e., the
component form) $l_{d}^{\mu }$ and $l_{b}^{\mu }$ which describe the
corresponding processes of light propagation (in the direct and opposite
directions).'' Then $,$in $S,$ the rest frame of the rod, $l_{d}^{\mu
}=(cL_{0}/c,L_{0},0,0)$ and $l_{b}^{\mu }=(cL_{0}/c,-L_{0},0,0),$ while in $%
S^{\prime },$ where the rod is moving, they are $l_{d}^{\mu ^{\prime
}}=(c\gamma L_{0}(1+\beta )/c),\gamma L_{0}(1+\beta ),0,0),$ and $l_{b}^{\mu
^{\prime }}=(c\gamma L_{0}(1-\beta )/c),-\gamma L_{0}(1-\beta ),0,0).$
Thence in $S$ one finds $l_{rel}^{\mu }=(l_{d}^{\mu }-l_{b}^{\mu
})/2=(0,L_{0},0,0)$ and in $S^{\prime }$ the component form of this 4-vector
of relativistic length is $l_{rel}^{\mu ^{\prime }}=(\gamma \beta
L_{0},\gamma L_{0},0,0).$ Now Strel'tsov, in the similar way as in the
asynchronous definition, compares only the spatial parts of $l_{rel}^{\mu
^{\prime }}$ and $l_{rel}^{\mu }$ and defines that the relativistic length
in $S^{\prime }$ is $l_{rel}^{\prime }\equiv l_{rel}^{1^{\prime }},$ which
is related with $l_{rel}\equiv l_{rel}^{1}$ in $S$ by the ''elongation
formula'' $l_{rel}^{\prime }=\gamma l_{rel}.$ These quantities $%
l_{rel}^{1^{\prime }}$ and $l_{rel}^{1}$ are considered to be the same
quantity for observers in $S^{\prime }$ and in $S.$ It is argued in \cite
{strel} that such ''approach has a manifestly relativistic covariant
character.'' But, as already said, the formulation of SR with true tensors
(or the CBGQs), i.e., the ''TT relativity,'' shows that comparison of only
spatial (or temporal) parts of the components of the distance 4-vector $%
l_{AB}^{a}$ in $S$ and $S^{\prime }$ is physically meaningless. Thus $%
l_{rel}^{1^{\prime }}$ and $l_{rel}^{1}$ are not the same quantity for
observers in $S^{\prime }$ and in $S.$ In general, as can be concluded from
the preceding sections, the spatial or temporal distances are not well
defined physical quantities in 4D spacetime. Consequently the radar
formulation of SR, together with the asynchronous formulation and Einstein's
formulation of SR, belongs to the ''AT\ relativity.'' Having discussed
different theoretical formulations of SR we can go to the comparison with
experiments.

\section{THE\ COMPARISON\ WITH\ EXPERIMENTS}

In numerous papers and textbooks it is considered that the experiments on
''length contraction'' and ''time dilatation'' test SR, but the discussion
from the previous sections shows that such an interpretation of the
experiments refers exclusively to the ''AT relativity,'' and not to the ''TT
relativity.'' We have shown that when SR is understood as the theory of 4D
spacetime with pseudo-Euclidean geometry then instead of the Lorentz
contraction and the dilatation of time one has to consider the 4D tensor
quantities, the spacetime length $l$ (\ref{elspat}), or the distance
4-vector $l_{AB}^{a}=x_{B}^{a}-x_{A}^{a}.$ Namely in the ''TT relativity''
the measurements in different IFRs (and different coordinatizations) have to
refer to the same 4D tensor quantity, i.e., to a CBGQ, (of course the same
holds for the theory). In the chosen IFR and the chosen coordinatization
(this choice defines what are the basis 4-vectors and 1-forms) the
measurement of some 4D quantity has to contain the measurements of all parts
(all the basis components) of such a quantity. However in almost all
experiments that refer to SR only the quantities belonging to the ''AT
relativity'' were measured. From the ''TT relativity'' viewpoint such
measurements are incomplete, since only some parts of a 4D quantity, not
all, are measured. This fact presents a serious difficulty in the reliable
comparison of the existing experiments with the ''TT relativity,'' and,
actually, we shall be able to compare in a quantitative manner only some of
the existing experiments with the ''TT relativity.'' This will be examined
in the comparison of the theoretical results for the spacetime length in the
''TT relativity'' and the spatial and temporal distances in the ''AT\
relativity'' with the existing experiments (see also \cite{ivtruII}). We
note that different test theories of SR have been proposed (see, e.g., \cite
{ander} and references therein), but ultimately all of them use the time
dilatation and length contraction parameters. (For example, even in the
recent test theory \cite{mans} which poses the question \cite{mans}: ''..
how accurately the background spacetime of physical phenomena, at least
locally, is a Minkowski spacetime?'' the authors states in the abstract:
''It is shown that the time dilatation and length contraction parameters
measure the deviation from a Riemannian geometry.'' Thence all of the
existing test theories are not actually test theories of \emph{SR}, but test
theories of the usual ''AT relativity'' approach to SR. Our aim in the
following sections, which deal with the comparison with experiments, is not
the comparison of some test theories with experiments, but the comparison of
the existing experimental results with different theoretical approaches to
SR, i.e., with the usual ''AT relativity'' and the ''TT relativity.'' It
will be shown that the ''TT relativity'' theoretical results agree with all
experiments that are complete from the ''TT relativity'' viewpoint, i.e., in
which all parts of the considered tensor quantity are measured in the
experiment. However the ''AT relativity'' results agree only with some of
the examined experiments and this agreement will exist only for the specific
coordinatization, i.e., the ''e'' coordinatization.

\section{THE ''MUON'' EXPERIMENT}

First we shall examine an experiment in which different results will be
predicted for different synchronizations in the conventional approach to SR,
i.e., in the ''AT relativity,'' but of course the same results for all
synchronizations will be obtained in the ''TT relativity.'' This is the
''muon'' experiment, which is theoretically discussed in Secs. 2.1 and 2.2$.$
The ''muon'' experiment is quoted in almost every textbook on general
physics, see, e.g., \cite{Feyn} and \cite{Kittel}. Moreover, an experiment 
\cite{Frisch} was the basis for a film often shown in introductory modern
physics courses: ''Time dilation: An experiment with $\mu $ mesons.''

In these experiments \cite{Frisch} (see also \cite{Easwar}) the fluxes of
muons on a mountain, $N_{m}$, and at sea level, $N_{s}$, are measured, and
the number of muons which decayed in flight is determined from their
difference. Also the distribution of the decay times is measured for the
case when the muons are at rest, giving a lifetime $\tau $ of approximately $%
2.2\mu s.$ The rate of decay of muons at rest, i.e., in the muon frame, is
compared with their rate of decay in flight, i.e., in the Earth frame. In 
\cite{Frisch} high-velocity muons are used, which causes that the fractional
energy loss of the muons in the atmosphere is negligible, making it a
constant velocity problem. The discussion of the ''muon'' experiment in
Secs. 2.1 and 2.2 referred to the decay of only one particle. When the real
experiments are considered, as in \cite{Frisch}, then we use data on the
decay of many such radioactive particles and the characteristic quantities
are avareged over many single decay events.

\subsection{The nonrelativistic approach}

In the nonrelativistic theory the space and time are separated. The
coordinate transformations connecting the Earth frame and the muon frame are
the Galilean transformations giving that $t_{E}$, the travel time from the
mountain to sea level when measured in the Earth frame, is the same as $%
t_{\mu }$, which is the elapsed time for the same travelling but measured in
the moving frame of the muon, $t_{E}=t_{\mu }$. Also, in the nonrelativistic
theory, the lifetimes of muons in the mentioned two frames are equal, $\tau
_{E}=\tau _{\mu }=\tau .$ Muon counts on the mountain $N_{m},$ and at sea
level $N_{s},$ as experimentally determined numbers, do not depend on the
frame in which they are measured and on the chosen coordinatization. This
result, i.e., that $N_{s\mu }$=$N_{sE}=N_{s}$ and $N_{m\mu }=N_{mE}=N_{m},$
has to be obtained not only in the nonrelativistic theory but also in the
''AT relativity'' and in the ''TT relativity.'' The differential equation
for the radioctive-decay processes in the nonrelativistic theory can be
written as 
\begin{equation}
dN/dt=-\lambda N,\quad N_{s}=N_{m}\exp (-t/\tau ).  \label{radclas}
\end{equation}
The travel time $t_{E}$ is not directly measured by clocks, but, in the
Earth frame, it is determined as the ratio of the height of the mountain $%
H_{E}$ and the velocity of the muons $v$, $t_{E}=H_{E}/v.$ The equation (\ref
{radclas}) holds in the Earth frame and in the muon frame too, since the two
frames are connected by the Galilean transformations, and, as mentioned
above, the corresponding times are equal, $t_{E}=t_{\mu }$ and $\tau
_{E}=\tau _{\mu }.$ Hence we conclude that in the nonrelativistic theory the
exponential factors are the same in both frames and consequently the
corresponding fluxes in the two frames are equal, $N_{s\mu }$=$N_{sE}$ and $%
N_{m\mu }=N_{mE}$, as it must be. However the experiments show that the
actual flux at sea level is much higher than that expected from such a
nonrelativistic calculation, and thus the nonrelativistic theory does not
agree with the experimental results.

\subsection{The usual ''AT relativity'' approach}

In the ''AT relativity'' different physical phenomena in different IFRs must
be invoked to explain the measured values of the fluxes; the time
''dilatation'' is used in the Earth frame, but in the muon frame one
explains the data by means of the Lorentz ''contraction.'' In order to
exploit the results of Secs. 2.1 and 2.2 we analyse the ''muon'' experiment
not only in the ''e'' coordinatization but also in the ''r''
coordinatization. As shown in Sec. 2.2 the ''AT relativity'' considers that
the spatial and temporal parts of the spacetime length are well-defined
physical quantities in 4D spacetime.

Then, as in the nonrelativistic theory, the equation for the
radioactive-decay in the ''AT relativity'' can be written as 
\begin{equation}
dN/dx^{0}=-\lambda N,\quad N_{s}=N_{m}\exp (-\lambda x^{0}).  \label{radate}
\end{equation}
The equation (\ref{radate}) contains a specific coordinate, the $x^{0}$
coordinate, which means that the equation (\ref{radate}) will not remain
unchanged upon the Lorentz transformation, i.e., it will not have the same
form in different IFRs (and also in different coordinatizations). But in the
''AT relativity'' it is not required that the physical quantities must be
the 4D tensor quantities that correctly transform upon the Lorentz
transformations. Thus the quantities in (\ref{radate}) are not the 4D tensor
quantities, i.e., they are not the true tensors or the CBGQs. This will
cause that different phenomena in different IFRs will need to be invoked to
explain the same physical effect, i.e., the same experimental data. In the
Earth frame and in the ''e'' coordinatization we can write in (\ref{radate})
that $x_{E}^{0}=ct_{E},$ $\lambda _{E}=1/c\tau _{E},$ which gives that the
radioactive-decay law becomes $N_{sE}=N_{mE}\exp (-t_{E}/\tau _{E}).$ In the
experiments \cite{Frisch} $N_{sE},$ $N_{mE},$ and $t_{E}=H_{E}/v$ are
measured in the Earth frame (tacitly assuming the ''e'' coordinatization).
However the lifetime of muons is measured in their rest frame. Now, in
contrast to the nonrelativistic theory where $\tau _{E}=\tau _{\mu }$ and $%
t_{E}=t_{\mu },$ the ''AT relativity'' assumes that in the ''e''
coordinatization there is the time ''dilatation'' determined by (\ref{tidil}%
), which gives the connection between the lifetimes of muons in the Earth
frame $\tau _{E}$ and the measured lifetime in the muon frame $\tau _{\mu }$
as 
\begin{equation}
\tau _{E}=\gamma \tau _{\mu }.  \label{taue}
\end{equation}
Using that relation one finds that the radioactive-decay law, when expressed
in terms of the measured quantities, becomes 
\begin{equation}
N_{sE}=N_{mE}\exp (-t_{E}/\tau _{E})=N_{mE}\exp (-t_{E}/\gamma \tau _{\mu }).
\label{radAT1}
\end{equation}
This equation is used in \cite{Frisch} to make the ''relativistic''
calculation and compare it with the experimental data. In fact, in \cite
{Frisch}, the comparison is made between the predicted time dilatation
factor $\gamma $ of the muons and an observed $\gamma .$ The predicted $%
\gamma $ is $8.4\pm 2,$ while the observed $\gamma $ is found to be $8.8\pm
0.8$, which is a convincing agreement. The prediction of $\gamma $ is made
from the measured energies of muons on the mountain and at sea level; these
energies are determined from the measured amount of material which muons
penetrated when stopped, and then the energies are converted to the speeds
of the muons using the relativistic relation between the total energy and
the speed. The observed $\gamma $ is determined from the relation (\ref
{radAT1}), where the measured rates were $N_{sE}=397\pm 9$ and $%
N_{mE}=550\pm 10,$ and the measured height of the mountain is $H_{E}=1907m.$
The lifetime of muons $\tau _{\mu }$ in the muon frame is taken as the
information from other experiments (in order to obtain more accurate result)
and it is $\tau _{\mu }=2.211\cdot 10^{-6}s.$

Let us now see how the experiments are interpreted in the muon frame. (We
note that \cite{Frisch} compared the theory (the ''AT relativity'') and the
experiments only in the Earth frame, but using $\tau _{\mu }$ from the muon
frame.) First we have to find the form of the law for the radioactive-decay
processes (\ref{radate}) in the muon frame. As considered above the
radioactive-decay law $N_{sE}=N_{mE}\exp (-t_{E}/\tau _{E})$ in the Earth
frame and in the ''e'' coordinatization is obtained from the equation (\ref
{radate}) using the relations $x_{E}^{0}=ct_{E}$ and $\lambda _{E}=1/c\tau
_{E}.$ But, as already said, the equation (\ref{radate}) does not remain
unchanged upon the Lorentz transformation. Accordingly it cannot have the
same form in the Earth frame and in the muon frame. So, actually, in the 4D
spacetime, the equation for the radioactive-decay processes in the muon
frame could have, in principle, a different functional form than the
equation (\ref{radAT1}), which describes the same radioactive- decay
processes in the Earth frame. However, in the ''AT relativity,'' despite of
the fact that the quantities in the Earth frame and in the muon frame are
not connected by the Lorentz transformations, the equation for the
radioactive-decay processes in the muon frame is obtained from the equation (%
\ref{radate}) in the same way as in the Earth frame, i.e., writting that $%
x_{\mu }^{0}=ct_{\mu },$ and $\lambda _{\mu }=1/c\tau _{\mu },$ whence 
\begin{equation}
N_{s\mu }=N_{m\mu }\exp (-t_{\mu }/\tau _{\mu }).  \label{radmu}
\end{equation}
The justification for such a procedure can be done in the following way. In
the ''AT relativity'' the principle of relativity acts as some sort of
''Deus ex machina,'' which resolves problems; the relation (\ref{radate}) is 
\emph{proclaimed} to be the physical law and the principle of relativity
requires that a physical law must have the same form in different IFRs.
(This is the usual way in which the principle of relativity is understood in
the ''AT relativity.'') Therefore, one can write in the equation (\ref
{radate}) that $x_{E}^{0}=ct_{E}$ and $\lambda _{E}=1/c\tau _{E}$ in the
Earth frame and $x_{\mu }^{0}=ct_{\mu },$ and $\lambda _{\mu }=1/c\tau _{\mu
}$ in the muon frame. With such substitutions the form of the law is the
same in both frames, as it is required by the principle of relativity. Then,
as we have already seen, when the consideration is done in the Earth frame,
the relation (\ref{taue}) for the time dilatation is used to connect
quantities in two frames, instead of to connect them by the Lorentz
transformations. When the consideration is performed in the muon frame
another relation is invoked to connect quantities in two frames. Namely it
is considered in the ''AT relativity'' that in the muon frame the mountain
is moving and the muon ''sees'' the height of the mountain Lorentz
contracted, 
\begin{equation}
H_{\mu }=H_{E}/\gamma ,  \label{hacon}
\end{equation}
which is Eq. (\ref{apcon}) for the Lorentz contraction, giving that 
\begin{equation}
t_{\mu }=H_{\mu }/v=H_{E}/\gamma v=t_{E}/\gamma .  \label{hami}
\end{equation}
This leads to the same exponential factor in (\ref{radmu}) as that one in
the Earth frame in (\ref{radAT1}), $\exp (-t_{\mu }/\tau _{\mu })=\exp
(-t_{E}/(\gamma \tau _{\mu })).$ From that result it is concluded that in
the ''AT relativity'' and in the ''e'' coordinatization the corresponding
fluxes are equal in the two frames, $N_{s\mu }$=$N_{sE}=N_{s}$ and $N_{m\mu
}=N_{mE}=N_{m}.$ Strictly speaking, it is not the mentioned equality of
fluxes, but the equality of ratios of fluxes, $N_{sE}/N_{mE}=$ $N_{s\mu
}/N_{m\mu }$, which follows from the equality of the exponential factors in (%
\ref{radAT1}) and (\ref{radmu}). In \cite{Frisch} the time $t_{\mu }$ that
the muons spent in flight according to their own clocks was inferred from
the measured distribution of decay times of muons at rest. Since the
predicted fluxes $N_{sE}$ and $N_{mE}$ are in a satisfactory agreement with
the measured ones, and since the theory (which deals with the time
dilatation and the Lorentz contraction) predicts their independence on the
chosen frame, it is generally accepted that the ''AT relativity'' correctly
explains the measured data.

The above comparison is worked out only in the ''e'' coordinatization, but
the physics demands that the independence of the fluxes on the chosen frame
must hold in all permissible coordinatizations. Therefore we now discuss the
experiments \cite{Frisch} from the point of view of the ''AT relativity''
but in the ''r'' coordinatization. Then, using (\ref{radate}), we can write
the relation for the fluxes in the ''r'' coordinatization and in the Earth
frame as 
\[
N_{r,sE}=N_{r,mE}\exp (-\lambda _{r,E}x_{r,E}^{0})=N_{r,mE}\exp
(-x_{r,E}^{0}/x_{r,E}^{0}(\tau _{E})), 
\]
where $x_{r,E}^{0}(\tau _{E})=1/\lambda _{r,E}.$ Again, as in the ''e''
coordinatization, we have to express $x_{r,E}^{0}(\tau _{E})$ in the Earth
frame in terms of the measured quantity $x_{r,\mu }^{0}(\tau _{\mu })$ using
the relation (\ref{tider}) for the time ''dilatation'' in the ''r''
coordinatization, 
\[
x_{r,E}^{0}(\tau _{E})=(1+2\beta _{r})^{1/2}c\tau _{\mu }. 
\]
Hence, the radioactive-decay law (\ref{radate}), in the ''r''
coordinatization, and when expressed in terms of the measured quantities,
becomes 
\begin{equation}
N_{r,sE}=N_{r,mE}\exp (-x_{r,E}^{0}/(1+2\beta _{r})^{1/2}c\tau _{\mu }),
\label{radiner}
\end{equation}
and it corresponds to the relation (\ref{radAT1}) in the ''e''
coordinatization. If we express $\beta _{r}$ in terms of $\beta =v/c$ as $%
\beta _{r}=\beta /(1-\beta )$ (see (\ref{elr})) and use (\ref{temini}) to
connect the ''r'' and ''e'' coordinatizations, $%
x_{r,E}^{0}=x_{E}^{0}-x_{E}^{1}=ct_{E}-H_{E},$ then the exponential factor
in (\ref{radiner}) becomes $=\exp \left\{ -(ct_{E}-H_{E})/\left[ (1+\beta
)/(1-\beta )\right] ^{1/2}c\tau _{\mu }\right\} .$ Using $H_{E}=vt_{E}$ this
exponential factor can be written in the form that resembles to that one in (%
\ref{radAT1}), i.e., it is $=\exp (-t_{E}/\Gamma _{rE}\tau _{\mu }),$ and (%
\ref{radiner}) can be written as 
\begin{equation}
N_{r,sE}=N_{r,mE}\exp (-t_{E}/\Gamma _{rE}\tau _{\mu }).  \label{raderAT}
\end{equation}
We see that $\gamma =(1-\beta )^{-1/2}$ in (\ref{radAT1}) (the ''e''
coordinatization) is replaced by a different factor 
\begin{equation}
\Gamma _{rE}=(1+\beta )^{1/2}(1-\beta )^{-3/2}=(1+\beta )(1-\beta
)^{-1}\gamma  \label{gama}
\end{equation}
in (\ref{raderAT}) (the ''r'' coordinatization). The \emph{observed} $\Gamma
_{rE}$ in the experiments \cite{Frisch} must remain the same, the observed $%
\Gamma _{rE}=8.8\pm 0.8,$ (it is determined from (\ref{raderAT}) with the
measured values of $N_{r,sE},\ N_{r,mE},\ t_{E}$ and $\tau _{\mu }$), but
the \emph{predicted} $\Gamma _{rE},$ using the above relation for $\Gamma
_{r}$ and the known, predicted, $\gamma =8.4\pm 2,$ becomes $\simeq
250\gamma ,$%
\begin{equation}
\Gamma _{rE}\simeq 250\gamma .  \label{gamaer}
\end{equation}
We see that from the common point of view a quite unexpected result is
obtained in the ''r'' coordinatization; the \emph{observed} $\Gamma _{rE}$
is as before $=8.8,$ while the \emph{predicted} $\Gamma _{rE}$ is $\simeq
250\cdot 8.4=2100.$ Similarly, one can show that there is a great
discrepancy between the fluxes measured in \cite{Frisch} and the fluxes
predicted when the ''dilatation'' of time is taken into account but in the
''r'' coordinatization and all is in the Earth frame. Furthermore, it can be
easily proved that predicted values in the ''r'' coordinatization and in the
muon frame will again greatly differ from the measured ones. \emph{Such
results explicitly show that the ''AT relativity'' is not a satisfactory
relativistic theory; it predicts, e.g., different values of the flux }$N_{s}$
\emph{(for the same measured} $N_{m}$\emph{)} \emph{in different
synchronizations and for some synchronizations these predicted values are
quite different but the measured ones. }These results are directly contrary
to the generally accepted opinion about the validity of the ''AT
relativity.''

\subsection{The ''TT relativity'' approach}

Let us now examine the experiments \cite{Frisch} from the point of view of
the ''TT relativity.'' In the ''TT relativity'' all quantities entering into
physical laws must be 4D tensor quantities, and thus with correct
transformation properties; \emph{the same 4D quantity} has to be considered
in different IFRs and different coordinatizations. In the usual, ''AT
relativity,'' analysis of the ''muon'' experiment, for example, the
lifetimes $\tau _{E}$ and $\tau _{\mu }$ are considered as the same
quantity. Although the transformation connecting $\tau _{E}$ and $\tau _{\mu
}$ (the dilatation of time (\ref{taue})) is only \emph{a part} of the
Lorentz transformation written in the ''e'' coordinatization, it is believed
by all proponents of the ''AT relativity'' that $\tau _{E}$ and $\tau _{\mu
} $ refer to the same temporal distance (the same quantity) but measured by
the observers in two relatively moving IFRs. However, as shown in the
preceding sections and in \cite{tom1} (see Fig.4), in 4D spacetime $\tau
_{E} $ and $\tau _{\mu }$ refer to different quantities, which are not
connected by the Lorentz transformation. To paraphrase Gamba \cite{gamba}:
''As far as relativity is concerned, quantities like $\tau _{E}$ and $\tau
_{\mu }$ are different quantities, not necessarily related to one another.
To ask the relation between $\tau _{E}$ and $\tau _{\mu }$ from the point of
view of relativity, is like asking what is the relation between the
measurement of the radius of the Earth made by an observer $S$ and the
measurement of the radius of Venus made by an observer $S^{\prime }.$ We can
certainly take the ratio of the two measures; what is wrong is the tacit
assumption that relativity has something to do with the problem just because
the measurements were made by \emph{two} observers.''

Hence, in the ''TT relativity,'' instead of the equation (\ref{radate}),
which explicitly contains only the specific coordinate, $x^{0}$ coordinate,
we formulate the radioactive-decay law in terms of true tensor quantities,
i.e., the CBGQs, as 
\begin{equation}
dN/dl=-\lambda N,\quad N=N_{0}\exp (-\lambda l).  \label{radTT}
\end{equation}
$l$ is the spacetime length defined by (\ref{elspat}), where $l^{a}(l^{b})$
is the distance 4-vector between two events $A$ and $B$, $%
l^{a}=l_{AB}^{a}=x_{B}^{a}-x_{A}^{a}$. $x_{A,B}^{a}$ are the position
4-vectors for the events of creation of muons (here on the mountain; we
denote it as the event $O$) and their arrival (here at sea level; the event $%
A$). $\lambda =1/l(\tau );$ $l(\tau )$ is the spacetime length for the
events of creation of muons (here on the mountain; the event $O$) and their
decay after the lifetime $\tau ,$ the event $T$. $l,$ defined in such a way,
is a geometrical quantity. Then in the ''e'' coordinatization and in the
muon frame the distance 4-vector $l_{OA}^{a},$ when written as the CBGQ,
becomes $l_{\mu ,OA}^{a}=ct_{\mu }e_{0}+0e_{1}$ (the subscript $\mu $ will
be used, as previously in this section, to denote the quantities in the muon
frame, while Greek indices $\alpha ,\beta $ denote the components of some
geometric object, e.g., the components $l_{\mu ,OA}^{\alpha }$ in the muon
frame of the distance 4-vector $l_{OA}^{a}$), and the spacetime length $l$
between these events is $l_{OA}=(l_{\mu ,OA}^{\beta }l_{\mu ,\beta
OA})^{1/2}=(-c^{2}t_{\mu }^{2})^{1/2}.$ The distance 4-vector $l_{OT}^{a}$
written as the CBGQ in the ''e'' coordinatization and in the muon frame is $%
l_{\mu ,OT}^{a}=c\tau _{\mu }e_{0}+0e_{1},$ whence the spacetime length $%
l_{OT}=(l_{\mu ,OT}^{\beta }l_{\mu ,\beta OT})^{1/2}=(-c^{2}\tau _{\mu
}^{2})^{1/2}.$ Inserting the spacetime lengths $l_{OA}$ and $l_{OT}$ into
the equation (\ref{radTT}) we find the expression for the radioactive-decay
law in the ''TT relativity'' 
\begin{equation}
N_{s}=N_{m}\exp (-l_{OA}/l_{OT}),  \label{radlaw}
\end{equation}
which in the ''e'' coordinatization and in the muon frame takes the same
form as the relation (\ref{radmu}) (the radioactive-decay law in the ''AT
relativity'' in the ''e'' coordinatization and in the muon frame), 
\begin{equation}
N_{s}=N_{m}\exp (-l_{OA}/l_{OT})=N_{m}\exp (-t_{\mu }/\tau _{\mu }).
\label{radla1}
\end{equation}
Since the spacetime length $l$ is independent on the chosen IFR and on the
chosen coordinatization \emph{the relation (\ref{radlaw}) holds in the same
form in the Earth frame and in the muon frame and in both coordinatizations,
the ''e'' and ''r'' coordinatizations.} Hence we do not need to examine Eq. (%
\ref{radlaw}) in the Earth frame, and in the ''r'' coordinatization, but we
can simply compare the relation (\ref{radla1}) with the experiments. (The
relation (\ref{comu}) gives the distance 4-vectors $l_{OA}^{a}$ and $%
l_{OT}^{a}$ written as the CBGQs in the ''e'' coordinatization in the muon
frame (the $S$ frame) and in the Earth frame (the $S^{\prime }$ frame) and
similarly happens with Eq. (\ref{coer}) in the ''r'' coordinatization.)

Thus we conclude that, in order to check the validity of the ''TT
relativity'' in the ''muon'' experiment, we would need, strictly speaking,
to measure, e.g., the lifetime $\tau _{\mu }$ and the time $t_{\mu }$ in the
muon frame, where they determine $l_{OT}$ and $l_{OA}$ respectively, and
then to measure \emph{the same events} (that determined $\tau _{\mu }$ and $%
t_{\mu }$ in the muon frame) in an IFR that is in uniform motion relative to
the muon frame (at us it is the Earth frame). Of course it is not possible
to do so in the real ''muon'' experiment but, nevertheless, in this case we
can use the data from experiments \cite{Frisch} and interpret them as that
they were obtained in the way required by the ''TT relativity.'' The reasons
for such a conclusion are the identity of microparticles of the same sort,
the assumed homogeneity and isotropy of the spacetime, and some other
reasons that are actually discussed in \cite{Frisch} (although from another
point of view). Here we shall not discuss this, in principle, a very complex
question, than we take the measured values of $\tau _{\mu },$ $t_{\mu },$ $%
N_{s}$ and $N_{m}$ and compare them with the results predicted by the
relation (\ref{radla1}). In \cite{Frisch} $\tau _{\mu }$ is taken to be $%
\tau _{\mu }=2.211\mu s,$ $N_{s}=397\pm 9,$ $N_{m}=550\pm 10,$ but $t_{\mu }$
is not measured than it is estimated from Fig. 6(a) in \cite{Frisch} to be $%
t_{\mu }=0.7\mu s.$ Inserting the values of $\tau _{\mu },$ $t_{\mu }$ and $%
N_{m}$ from \cite{Frisch} (for this simple comparison we take only the mean
values without errors) into (\ref{radla1}) we predict that $N_{s}$ is $%
N_{s}=401,$ which is in an excellent agreement with the measured $N_{s}=397.$
As it is already said, the spacetime length $l$ takes the same value in both
frames and both coordinatizations, $l_{e,\mu }=l_{e,E}=l_{r,\mu }=l_{r,E}.$
Hence, for the measured $N_{m}=550$ and if the distance 4-vectors $%
l_{OA}^{a} $ and $l_{OT}^{a}$ would be measured in the Earth frame, and in
both frames in the ''r'' coordinatization, we would find the same $%
N_{s}=401. $ This result undoubtedly confirms the consistency and the
validity of the ''TT relativity.''

\emph{The nonrelativistic theory predicts the same value of the exponential
factor in both frames, }$\exp (-t_{E}/\tau _{E})=\exp (-t_{\mu }/\tau _{\mu
}),$\emph{\ since it deals with the absolute time, i.e., with the Galilean
transformations. But, for the measured }$N_{m}$\emph{\ the nonrelativistic
theory predicts too small }$N_{s}.$\emph{\ The ''AT relativity'' correctly
predicts the value of }$N_{s}$\emph{\ in both frames but only in the ''e''
coordinatization, while in the ''r'' coordinatization the experimental }$%
N_{s}$\emph{\ and the theoretically predicted }$N_{s}$\emph{\ drastically
differ. The ''TT relativity'' completely agrees with the experiments in all
IFRs and all permissible coordinatizations. Thus, the ''TT relativity,'' as
the theory of 4D spacetime with the pseudo-Euclidean geometry, is in a
complete agreement with the experiments.}

\subsection{Another time ''dilatation'' experiments}

The same conclusion can be achieved comparing the other particle lifetime
measurements, e.g., \cite{rossi}, or for the pion lifetime \cite{ayres},
with all three theories. However, as it is already said, all the mentioned
experiments, and not only them but all other too, were designed to test the
''AT relativity.'' Thus in the experiments \cite{rossi}, which preceded to
the experiments \cite{Frisch} and \cite{Easwar}, the relation similar to (%
\ref{radAT1}) is used but with $t_{E}$ replaced by $H_{E}$ (=$vt_{E}$) and $%
\tau _{E}$ (the lifetime of muons in the Earth frame) replaced by $L$ $%
=v\tau _{E}$ ($L$ is the ''average range before decay''), and also the
connection between the lifetimes (\ref{taue}) ($\tau _{E}=\gamma \tau _{\mu
} $) is employed. Obviously the \emph{predictions} of the results in the
experiments \cite{rossi} will depend on the chosen synchronization, since
they deal with the ''AT relativity'' and use the radioactive-decay law in
the form that contains only a part of the distance 4-vector. The \emph{%
predictions }obtained by the use of the ''TT relativity'' will be again
independent on the chosen IFR and the chosen coordinatization. However the
comparison of these experiments \cite{rossi} with the ''TT relativity'' is
difficult since, e.g., they have no data for $t_{\mu }.$ Similarly happens
with the experiments reported in \cite{ayres}.

The lifetime measurements of muons in the g-2 experiments \cite{bailey} are
often quoted as the most convincing evidence for the time dilatation, i.e.,
they are claimed as high-precision evidence for SR. Namely in the literature
the evidence for the time dilatation is commonly considered as the evidence
for SR. The muon lifetime in flight $\tau $ is determined by fitting the
experimental decay electron time distribution to the six-parameter
phenomenological function describing the normal modulated exponential decay
spectrum (their Eq.(1)). Then by the use of the relation $\tau =\gamma \tau
_{0}$ and of $\tau _{0}$ (our $\tau _{\mu }$), the lifetime at rest (as
determined by other workers), they obtained the time-dilatation factor $%
\gamma ,$ or the kinematical $\gamma .$ This $\gamma $ is compared with the
corresponding dynamical $\gamma $ factor ($\gamma =(p/m)dp/dE$), which they
called $\overline{\gamma }$ (the average $\gamma $ value). $\overline{\gamma 
}$ is determined from the mean rotation frequency $\overline{f}_{rot}$ by
the use of the Lorentz force law (the ''relativistic'' expression); the
magnetic field was measured in terms of the proton NMR frequency $f_{p}$
(for the discussion of $g-2$ experiments within the traditional ''AT
relativity'' see also \cite{newman}). Limits of order $10^{-3}$ in $(\gamma -%
\overline{\gamma })/\gamma $ at the kinematical $\gamma =29.3$ were set. In
that way they also compared the value of the $\mu ^{+}$ lifetime at rest $%
\tau _{0}^{+}$ (from the other precise measurements) with the value found in
their experiment $\tau ^{+}/\overline{\gamma },$ and obtained $(\tau
_{0}^{+}-\tau ^{+}/\overline{\gamma })/\tau _{0}^{+}=(2\pm 9)\times 10^{-4},$
(this is the same comparison as the mentioned comparison of $\gamma $ with $%
\overline{\gamma }$). They claimed: ''At $95\%$ confidence the fractional
difference between $\tau _{0}^{+}$ and $\tau ^{+}/\overline{\gamma }$ is in
the range $(-1.6-2.0)\times 10^{-3}$.'' and ''To date, this is the most
accurate test of relativistic time dilation using elementary particles.''
The objections to the precision of the experiments \cite{bailey}, and the
remark that a convincing direct test of SR must not assume the validity of
SR in advance (in the use of the ''relativistic'' Lorentz force law in the
determination of the mean rotation frequency and thus of $\overline{\gamma }%
, $ and $\tau _{0}$), have been raised in \cite{huang}. The discussion of
these objections is given in \cite{field}.

However, our objections to \cite{bailey} are of a quite different nature.
Firstly, the theoretical relations refer to the ''e'' coordinatization and,
e.g., Eq.(1) in the first paper in \cite{bailey} cannot be transformed in an
appropriate way to the ''r'' coordinatization in order to compare the ''AT
relativity'' in different coordinatizations with the experiments. If only
the exponential factor is considered then this factor is again, as in \cite
{Frisch}, affected by synchrony choice. Although the time $t$ in that
exponential factor may be independent of the chosen synchronization (when $t$
is taken to be the multiple of the mean rotation period $T$), but $\tau $
does not refer to the events that happen at the same spatial point and thus
it is synchrony dependent quantity. This means that in the ''r''
coordinatization one cannot use the relation $\tau =\gamma \tau _{0}$ to
find the ''dilatation'' factor $\gamma ,$ but the relation (\ref{tider}) for
the time ''dilatation'' in the ''r'' coordinatization, $x_{r}^{0}(\tau
)=(1+2\beta _{r})^{1/2}c\tau _{0}$ must be employed. Hence, the whole
comparison of $\gamma $ with $\overline{\gamma }$ holds only in the ''e''
coordinatization; in another coordinatization the ''AT relativity'' predicts
quite different $\tau _{0}$ for the same $x^{0}(\tau )$ (that is inferred
from the exponential decay spectrum).

Let us now examine the measurements \cite{bailey} from the point of view of
the ''TT relativity.'' But for the ''TT relativity'' these experiments are
incomplete and cannot be compared with the theory. Namely, in the ''TT
relativity,'' as already said, it is not possible to find the values of the
muon lifetime in flight $\tau $ by analyses of the measurements of the
radioactive decay distribution, since, there, the radioactive decay law is
written in terms of the spacetime lengths and not with $t$ and $\tau .$
Also, in the ''TT relativity,'' there is not the connection between the muon
lifetime in flight $\tau $ and the lifetime at rest $\tau _{0}$ in the form $%
\tau =\gamma \tau _{0},$ since $\tau ,$ in the ''TT relativity,'' does not
exist as a well defined quantity. Thus, in the ''TT relativity,'' there is
no sense in the use of the relation $\tau =\gamma \tau _{0}$ to determine $%
\gamma .$ An important remark is in place here; in principle, in the ''TT
relativity,'' the same events and the same quantities have to be considered
in different frames of reference. This means that in the muon experiment 
\cite{bailey} the lifetime at rest $\tau _{0}$ refers to the decaying
particle in an accelerated frame and for the theoretical discussion we would
need to use the coordinate transformations connecting an IFR with an
accelerated frame of reference. (An example of the generalized Lorentz
transformation is given in \cite{nelson} but they are written in the ''e''
coordinatization and thus not in fully covariant way, i.e., not in the way
as we have written the covariant Lorentz transformation (\ref{fah}).)
Furthermore, in the experiments \cite{bailey} the average value of $\gamma $
($\overline{\gamma }$), i.e., the dynamical $\gamma ,$ for the circulating
muons is found by analysis of the bunch structure of the stored muon and the
use of the relation connecting $\overline{\gamma }$ and the mean rotation
frequency $\overline{f}_{rot}.$ This relation is obtained by the use of the
expression for the ''relativistic,'' i.e., the ''AT relativity,'' Lorentz
force law, which is expressed by means of the 3-vectors $\mathbf{E}$ and $%
\mathbf{B.}$ However, in contrast to the ''AT relativity,'' and also to the
usual covariant formulation, in the ''TT relativity,'' the Lorentz force as
the true tensor $K^{a}=(q/c)F^{ab}u_{b}$ ($F^{ab}$ is the electromagnetic
field tensor and $u^{b}$ is the 4-velocity of a charge $q$, see \cite{Wald}, 
\cite{vanzel} and \cite{tom1}) cannot be expressed in terms of the 3-vectors 
$\mathbf{E}$ and $\mathbf{B.}$ Namely in the ''AT relativity'' the real
physical meaning is attributed not to $F^{ab}$ but to the 3-vectors $\mathbf{%
E}$ and $\mathbf{B,}$ while in the ''TT relativity'' only the true tensor
quantities, or equivalently the CBGQs, do have well-defined physical meaning
both in the theory and in experiments. (The transformations of the 3-vectors 
$\mathbf{E}$\ and $\mathbf{B}$ are not directly connected with the Lorentz
transformations of the \emph{whole 4D tensor quantity} $F^{ab}$ as a
geometrical quantity$,$\ but indirectly through the transformations of \emph{%
some components} of $F^{ab},$\ and that happens \emph{in the specific
coordinatization, the Einstein coordinatization.} This issue is discussed in
detail in \cite{tom1}, where it is also shown that the 3-vector $\mathbf{E}$
($\mathbf{B}$) in an IFR $S$ and the transformed 3-vector $\mathbf{E}%
^{\prime }$ ($\mathbf{B}^{\prime }$) in relatively moving IFR $S^{\prime }$
do not refer to the same physical quantity in 4D spacetime, i.e., that the
conventional transformations of $\mathbf{E}$ and $\mathbf{B}$ are the AT.)
>From \cite{vanzel} and \cite{tom1} one can see how the Lorentz force $K^{a}$
is expressed in terms of the 4-vectors $E^{a}$ and $B^{a}$ and show when
this form corresponds to the classical expression for the Lorentz force with
the 3-vectors $\mathbf{E}$ and $\mathbf{B.}$ Also it can be seen from \cite
{tom1} and \cite{ivsc98} that for $B^{\alpha }\neq 0$ ($B^{\alpha }$ is the
component form of $B^{a}$ in the ''e'' coordinatization) it is not possible
to obtain $\gamma _{u}=1$ (the 4-velocity of a charge $q$ in the ''e''
coordinatization is $u^{\alpha }=(\gamma _{u}c,\gamma _{u}\mathbf{u})$ and $%
\gamma _{u}=(1-u^{2}/c^{2})^{-1/2}$), and the invariant Lorentz force $K^{a}$
can never take the form of the usual magnetic force $\mathbf{F}_{B}.$ Hence
it follows that in the ''TT relativity'' it is not possible to use the
Lorentz force $\mathbf{F}_{B}$ and the usual equation of motion $d(\overline{%
\gamma }m\mathbf{u)/}dt\mathbf{=}q(\mathbf{u}\times \mathbf{B})$ to find the
relation connecting $\overline{\gamma }$ and the mean rotation frequency $%
\overline{f}_{rot},$ and thus to find $\tau _{0}$ from $\tau /\overline{%
\gamma },.$in the way as in \cite{bailey}. The discussion about the
kinematical $\gamma $ (the relation $\tau =\gamma \tau _{0}$) and about the
dynamical $\overline{\gamma }$ (from the use of the Lorentz force) shows
that the measurements \cite{bailey} cannot be compared with the ''TT
relativity.'' But, as we explained before, in contrast to the usual opinion,
these experiments do not confirm the ''AT relativity'' either. Namely if the
exponential decay spectrum is analyzed in another coordinatization, e.g.,
the ''r'' coordinatization, then, similarly as for the experiments \cite
{Frisch}, one finds that for the given $N_{0}$ the theoretical and the
experimental $N$ differ.

\section{THE MICHELSON-MORLEY EXPERIMENT}

These conclusions will be further supported considering some other
experiments, which, customarily, were assumed to confirm the usual ''AT
relativity,'' that is, the Einstein formulation of SR. The first one will be
the famous Michelson-Morley experiment \cite{michels}, and some modern
versions of this experiment will be also discussed. Since the
Michelson-Morley experiment is considered in detail in \cite{tom2} we only
briefly discuss some results.

In the Michelson-Morley experiment two light beams emitted by one source are
sent, by half-silvered mirror $O$, in orthogonal directions. These partial
beams of light traverse the two equal (of the length $L$) and perpendicular
arms $OM_{1}$ (perpendicular to the motion) and $OM_{2}$ (in the line of
motion) of Michelson's inteferometer and the behaviour of the interference
fringes produced on bringing together these two beams after reflection on
the mirrors $M_{1}$ and $M_{2}$ is examined. In order to avoid the influence
of the effect that the two lengths of arms are not exactly equal the entire
inteferometer is rotated through $90^{0}.$ Then any small difference in
length becomes unimportant. The experiment consists of looking for a shift
of the intereference fringes as the apparatus is rotated. The expected
maximum shift in the number of fringes (the measured quantity) on a $90^{0}$%
\ rotation is 
\begin{equation}
\bigtriangleup N=\bigtriangleup (\phi _{2}-\phi _{1})/2\pi ,  \label{delfi}
\end{equation}
where $\bigtriangleup (\phi _{2}-\phi _{1})$ is the change in the phase
difference when the interferometer is rotated through $90^{0}.$ $\phi _{1}$
and $\phi _{2}$ are the phases of waves moving along the paths $OM_{1}O$ and 
$OM_{2}O,$ respectively.

\subsection{The ''TT relativity '' approach}

The Michelson-Morley experiment will be examined from the ''TT relativity''
viewpoint and then it will be shown how the usual ''AT relativity'' results
are obtained. The relevant quantity is the phase of a light wave, and it is
(when written in the abstract index notation) 
\begin{equation}
\phi =k^{a}g_{ab}l^{b},  \label{phase}
\end{equation}
where $k^{a}$ is the propagation 4-vector, $g_{ab}$ is the metric tensor and 
$l^{b}$ is the distance 4-vector. All quantities in (\ref{phase}) are true
tensor quantities. As discussed in Sec. 2 these quantities can be written in
the coordinate-based geometric language and, e.g., the decompositions of $%
k^{a}$ in $S$ and $S^{\prime }$ and in the ''e'' and ''r'' coordinatizations
are 
\begin{equation}
k^{a}=k^{\mu ^{\prime }}e_{\mu ^{\prime }}=k^{\mu }e_{\mu }=k_{r}^{\mu
^{\prime }}r_{\mu ^{\prime }}=k_{r}^{\mu }r_{\mu },  \label{kad}
\end{equation}
where the basis components $k^{\mu }$ of the CBGQ in the ''e''
coordinatization are transformed by $L^{\mu ^{\prime }}{}_{\nu ,e}$ (\ref
{lorus}), while the basis vectors $e_{\mu }$ are transformed by the inverse
transformation $(L^{\mu ^{\prime }}{}_{\nu ,e})^{-1}=L^{\mu }{}_{\nu
^{\prime },e}.$ Similarly holds for the ''r'' coordinatization where the
Lorentz transformation $L^{\mu ^{\prime }}{}_{\nu ,r}$ (\ref{elr}) has to be
used. By the same reasoning the phase $\phi $ (\ref{phase}) is given in the
coordinate-based geometric language as 
\begin{equation}
\phi =k_{e}^{\mu }g_{\mu \nu ,e}\,l_{e}^{\nu }=k_{e}^{\mu ^{\prime }}g_{\mu
\nu ,e}\,l_{e}^{\nu ^{\prime }}=k_{r}^{\mu }g_{\mu \nu ,r}\,l_{r}^{\nu
}=k_{r}^{\mu ^{\prime }}g_{\mu \nu ,r}\,l_{r}^{\nu ^{\prime }},  \label{pha2}
\end{equation}
(Note that the Lorentz transformation $L^{\mu ^{\prime }}{}_{\nu ,e}$ (\ref
{lorus}) and also $L^{\mu ^{\prime }}{}_{\nu ,r}$ (\ref{elr}) are the TT,
i.e., the isometries, and hence $g_{\mu \nu ,e}=g_{\mu ^{\prime }\nu
^{\prime },e}$, $g_{\mu \nu ,r}=g_{\mu ^{\prime }\nu ^{\prime },r}$, what is
already taken into account in (\ref{pha2}).) The traditional derivation of $%
\bigtriangleup N$ (see \cite{tom2} and, e.g., \cite{Feyn}, \cite{Kittel}, or
an often cited paper on modern tests of special relativity \cite{haugan})
deals only with the calculation of $t_{1}$ and $t_{2}$ in $S$ and $%
t_{1}^{\prime }$ and $t_{2}^{\prime }$ in $S^{\prime },$ but does not take
into account either the changes in frequencies due to the Doppler effect or
the aberration of light. (The Earth frame is the rest frame of the
interferometer, i.e.,\textbf{\ }it is the $S$\ frame, while the $S^{\prime }$%
\ frame is the (preferred) frame\ in which the interferometer is moving at
velocity $\mathbf{v.}$ In the $S$\ frame $t_{1}$ and $t_{2}$ are the times
required for the complete trips $OM_{1}O$ and $OM_{2}O$ respectively, while $%
t_{1}^{\prime }$ and $t_{2}^{\prime }$ are the corresponding times in $%
S^{\prime }.$) The ''AT relativity'' calculations \cite{drisc} and \cite
{schum} improve the traditional procedure taking into account the changes in
frequencies \cite{drisc} and the aberration of light \cite{schum}. But all
these approaches explain the experiments using the AT, the Lorentz
contraction and the time dilatation, and furthermore they always work only
in the ''e'' coordinatization. None of the ''AT relativity'' calculations
deal with the true tensors or with the CBGQs (comprising both components and
a basis). In this case such 4D tensor quantity is the phase (\ref{phase}) or
(\ref{pha2}). \emph{In the ''TT relativity'' approach to SR neither the
Doppler effect nor the aberration of light exist separately as well defined
physical phenomena.} \emph{The separate contributions to }$\phi $\emph{\ (%
\ref{phase}), or (\ref{pha2}), of the }$\omega t$\emph{\ (i.e., }$k^{0}l_{0}$%
\emph{)} \emph{factor\ \cite{drisc} and }$\mathbf{kl}$\emph{\ (i.e., }$%
k^{i}l_{i}$) \emph{factor\ \cite{schum} are, in general case, meaningless in
the ''TT relativity.'' From the ''TT relativity'' viewpoint only their
indivisible unity, the phase }$\phi $\ \emph{(\ref{phase}), or (\ref{pha2}),
is a correctly defined 4D quantity. }All quantities in (\ref{phase}), i.e., $%
k^{a}$, $g_{ab},$ $l^{b}$ and $\phi ,$ are the true tensor quantities, which
means that in all relatively moving IFRs and in all permissible
coordinatizations always \emph{the same 4D quantity}, e.g., $k^{a},$ or $%
l^{b},$ or $\phi ,$\ is considered. (Eq. (\ref{pha2}) shows it for $\phi $.)
This is not the case in the ''AT relativity.'' There, for example, the
relation for the time dilatation $t_{1}^{\prime }=\gamma t_{1},$ which is
used in the usual explanation (see, e.g., \cite{Feyn}, \cite{Kittel} and 
\cite{haugan}) of the Michelson-Morley experiment, is not the Lorentz
transformation of some 4D quantity, and $t_{1}^{\prime }$ and $t_{1}$ do not
correspond to the same 4D quantity considered in $S^{\prime }$ and $S$
respectively but to different 4D quantities, as can be clearly seen from
Sec. 2.2 (see \ref{tidil}). Only in the ''e'' coordinatization the $\omega t$
and $\mathbf{kl}$ factors can be considered separately. Therefore, and in
order to retain the similarity with the prerelativistic and the ''AT
relativity'' considerations, we first determine $\phi $ (\ref{phase}), (\ref
{pha2}), in the ''e'' coordinatization and in the $S$ frame (the rest frame
of the interferometer). This means that $\phi $ will be calculated from (\ref
{pha2}) as the CBGQ $\phi =k_{e}^{\mu }g_{\mu \nu ,e}\,l_{e}^{\nu }.$

Let now $A,$ $B$ and $A_{1}$ denote the events; the departure of the
transverse ray from the half-silvered mirror $O,$ the reflection of this ray
on the mirror $M_{1}$ and the arrival of this beam of light after the round
trip on the half-silvered mirror $O,$ respectively. In the same way we have,
for the longitudinal arm of the inteferometer, the corresponding events $A,$ 
$C$ and $A_{2}.$ To simplify the notation we omit the subscript 'e' in all
quantities. Then $k_{AB}^{\mu }$ and $l_{AB}^{\mu }$ (the basis components
of $k_{AB}^{a}$ and $l_{AB}^{a}$ in the ''e'' coordinatization and in $S$)
for the wave on the trip $OM_{1}$ (the events $A$ and $B$) are $k_{AB}^{\mu
}=(\omega /c,0,2\pi /\lambda ,0),$ $l_{AB}^{\mu }=(ct_{M_{1}},0,\overline{L}%
,0)$. For the wave on the return trip $M_{1}O,$ (the events $B$ and $A_{1}$) 
$k_{BA_{1}}^{\mu }=(\omega /c,0,-2\pi /\lambda ,0)$ and $l_{BA_{1}}^{\mu
}=(ct_{M_{1}},0,-\overline{L},0)$ (the elapsed times $t_{OM_{1}}$ and $%
t_{M_{1}O}$ for the trips $OM_{1}$ and $M_{1}O$ respectively are equal and
denoted as $t_{M_{1}}$, $t_{OM_{1}}=t_{M_{1}O}=t_{M_{1}}$). Hence the
increment of phase $\phi _{1}$ for the the round trip $OM_{1}O,$ is 
\begin{equation}
\phi _{1}=k_{AB}^{\mu }\,l_{\mu AB}+k_{BA_{1}}^{\mu }l_{\mu
BA_{1}}=2(-\omega t_{M_{1}}+(2\pi /\lambda )\overline{L}),  \label{fijd}
\end{equation}
where $\omega $ is the angular frequency. $L$ is the length of the segment $%
OM_{2}$ and $\overline{L}=L(1+\varepsilon )$ ($\varepsilon \ll 1$) is taken
to be, as in \cite{drisc}, the length of the arm $OM_{1}.$ As explained in 
\cite{drisc}: ''The difference $\overline{L}-L=\varepsilon L$ is usually a
few wavelengths ($\prec 25$) and is essential for obtaining useful
interference fringes.'' $L,$ $\overline{L}$ and $\nu $ are determined in $S$%
, the rest frame of the interferometer. Using the Lorentz transformation $%
L^{\mu ^{\prime }}{}_{\nu ,e}$ (\ref{lorus}) one can find $k^{\mu ^{\prime }}
$ and $l^{\mu ^{\prime }}$ in the ''e'' coordinatization and in $S^{\prime }$
for the same trips as in $S$. Then it can be easily shown that $\phi
_{1}^{\prime }$ in $S^{\prime }$ is the same as in $S,$ $\phi _{1}^{\prime
}=\phi _{1}.$ Also using the transformation matrix $T^{\mu }{}_{\nu ,r}$ (%
\ref{teer}), which transforms the ''e'' coordinatization to the ''r''
coordinatization, one can get all quantities in the ''r'' coordinatization
and in $S$, and then by the Lorentz transformation $L^{\mu ^{\prime
}}{}_{\nu ,r}$ (\ref{elr}) these quantities can be determined in the ''r''
coordinatization and in $S^{\prime }$. $\phi _{1}$ will be always the same
in accordance with (\ref{pha2}). Note that $g_{\mu \nu ,r}$ (\ref{geer})
from Sec. 2 has to be used in the calculation of $\phi $ in the ''r''
coordinatization. As an example we quote $k_{AB,r}^{\mu }$ and $%
l_{AB,r}^{\mu }$: $k_{AB,r}^{\mu }=((\omega /c)-2\pi /\lambda ,0,2\pi
/\lambda ,0)$ and $l_{AB,r}^{\mu }=(ct_{M_{1}}-\overline{L},0,\overline{L}%
,0).$ Hence, using $g_{\mu \nu ,r}$ one easily finds that 
\[
\phi _{AB,r}=k_{r}^{\mu }g_{\mu \nu ,r}\,l_{r}^{\nu }=(-\omega
t_{M_{1}}+(2\pi /\lambda )\overline{L})=\phi _{AB,e}.
\]
For further purposes we shall also need $k_{AB,r}^{\mu ^{\prime }}$ and $%
l_{AB,r}^{\mu ^{\prime }}.$ They are $k_{AB,r}^{\mu ^{\prime }}=((\gamma
\omega /c)(1+\beta )-2\pi /\lambda ,-\beta \gamma \omega /c,2\pi /\lambda ,0)
$ and $l_{AB,r}^{\mu ^{\prime }}=(\gamma ct_{M_{1}}(1+\beta )-\overline{L}%
,-\beta \gamma ct_{M_{1}},\overline{L},0)$ which yields 
\[
\phi _{AB,r}^{\prime }=\phi _{AB,r}=\phi _{AB,e}^{\prime }=\phi _{AB,e}.
\]
In a like manner we find $k_{AC}^{\mu }$ and $l_{AC}^{\mu }$ for the wave on
the trip $OM_{2},$ (the corresponding events are $A$ and $C$) as $%
k_{AC}^{\mu }=(\omega /c,2\pi /\lambda ,0,0)$ and $l_{AC}^{\mu
}=(ct_{M_{2}},L,0,0).$ For the wave on the return trip $M_{2}O$ (the
corresponding events are $C$ and $A_{2}$) $k_{CA_{2}}^{\mu }=(\omega
/c,-2\pi /\lambda ,0,0)$ and $l_{CA_{2}}^{\mu }=(ct_{M_{2}},-L,0,0)$) ($%
t_{OM_{2}}=t_{M_{2}O}=t_{M_{2}}$), whence 
\begin{equation}
\phi _{2}=k_{AC}^{\mu }\,l_{\mu AC}+k_{CA_{2}}^{\mu }l_{\mu
CA_{2}}=2(-\omega t_{M_{2}}+(2\pi /\lambda )L).  \label{f2jd}
\end{equation}
Of course one finds the same $\phi _{2}$ in $S$ and $S^{\prime }$ and in the
''e'' and ''r'' coordinatizations. Hence 
\begin{equation}
\phi _{1}-\phi _{2}=-2\omega (t_{M_{1}}-t_{M_{2}})+2(2\pi /\lambda )(%
\overline{L}-L).  \label{phasdif}
\end{equation}
Particularly for $\overline{L}=L,$ and consequently $t_{M_{1}}=t_{M_{2}},$
one finds $\phi _{1}-\phi _{2}=0.$ It can be easily shown that the same
difference of phase (\ref{phasdif}) is obtained in the case when the
interferometer is rotated through $90^{0},$ whence we find that $%
\bigtriangleup (\phi _{1}-\phi _{2})=0,$ and $\bigtriangleup N=0.$ \emph{%
According to the construction }$\phi $\emph{\ (\ref{phase}), or (\ref{pha2}%
), is a frame independent quantity and it also does not depend on the chosen
coordinatization in a considered IFR.} Thus we conclude that 
\begin{equation}
\bigtriangleup N_{e}=\bigtriangleup N_{e}^{\prime }=\bigtriangleup
N_{r}=\bigtriangleup N_{r}^{\prime }=0.  \label{delshif}
\end{equation}
This result is in a complete agreement with the Michelson-Morley \cite
{michels} experiment.

Driscoll \cite{drisc} improved the traditional ''AT relativity'' derivation
of the fringe shift taking into account the changes in frequencies due to
the Doppler effect. This improvement resulted in a ''surprising'' non-null
fringe shift 
\begin{equation}
\bigtriangleup N^{\prime }=\bigtriangleup (\phi _{2}^{\prime }-\phi
_{1}^{\prime })/2\pi =4(L\nu /c)\beta ^{2},  \label{shift}
\end{equation}
and we see that the entire fringe shift is due to the Doppler shift (see 
\cite{drisc} and \cite{tom2}). It is explicitly shown in \cite{tom2} that
Driscoll's result \emph{can be easily obtained from our ''TT relativity''
approach taking only the product }$k_{e}^{0^{\prime }}l_{0^{\prime }e}$\emph{%
\ in the calculation of the increment of phase }$\phi _{e}^{\prime }$\emph{\
in} $S^{\prime }$ in which the apparatus is moving.

We remark that the non-null fringe shift (\ref{shift}) would be quite
different in another coordinatization, e.g., in the ''r'' coordinatization,
since only a part $k_{e}^{0^{\prime }}l_{0^{\prime }e}$ of the whole 4D
tensor quantity $\phi $ (\ref{phase}) or (\ref{pha2}) is considered. Thus
when only a part of the whole phase $\phi $ (\ref{phase}) or (\ref{pha2}) is
taken into account then it leads to an unphysical result.

As shown in \cite{tom2} \emph{the same calculation of} $k^{i^{\prime
}}l_{i^{\prime }},$ \emph{the contribution of the spatial parts of} $k^{\mu
^{\prime }}$ \emph{and} $l_{\mu ^{\prime }}$ \emph{to} $\bigtriangleup
N_{e}^{\prime },$ \emph{shows that this term exactly cancel the} $%
k^{0^{\prime }}l_{0^{\prime }}$ \emph{contribution (Driscoll's non-null
fringe shift (\ref{shift})), yielding that} $\bigtriangleup N_{e}^{\prime
}=\bigtriangleup N_{e}=0.$ Thus the ''TT relativity''approach to SR
naturally explains the reason for the existence of Driscoll's non-null
fringe shift (\ref{shift}).

The results of the usual ''AT relativity'' calculation \emph{can be easily
explained from our true tensor formulation of SR taking only the part }$%
k_{e}^{0}l_{0^{\prime }e}$\ \emph{of the whole phase }$\phi $ \emph{(\ref
{phase}) or (\ref{pha2}) in the calculation of the increment of phase }$\phi
_{e}^{\prime }$\emph{\ in} $S^{\prime }.$ In contrast to Driscoll's
treatment the traditional analysis considers the part $k_{e}^{0}l_{0e}$ (of
the whole phase $\phi $ (\ref{phase}), (\ref{pha2})) in $S,$ the rest frame
of the interferometer, and $k_{e}^{0}l_{0^{\prime }e}$\ in $S^{\prime }$, in
which the apparatus is moving. $k_{e}^{0}$\emph{\ is not changed in
transition from} $S$\emph{\ to} $S^{\prime }$. Thus the increment of phase $%
\phi _{1}$ for the round trip $OM_{1}O$ in $S$, is 
\begin{equation}
\phi
_{1}=k_{AB}^{0}\,g_{00,e}l_{AB}^{0}+k_{BA_{1}}^{0}g_{00,e}l_{BA_{1}}^{0}=-2(%
\omega /c)(ct_{M_{1}})=-2\omega t_{M_{1}}.  \label{apfi1}
\end{equation}
In the $S^{\prime }$ frame we find for the same trip that 
\begin{equation}
\phi _{1}^{\prime }=k_{AB}^{0}\,l_{0^{\prime }AB}+k_{BA_{1}}^{0}l_{0^{\prime
}BA_{1}}=-2(\omega /c)(\gamma ct_{M_{1}})=-2\omega (\gamma t_{M_{1}}).
\label{apf1cr}
\end{equation}
This is exactly the result obtained in the traditional analysis (see \cite
{Feyn} or \cite{Kittel}) which is inerpreted as that there is a time
''dilatation'' $t_{1}^{\prime }=\gamma t_{1}$. In the same way we find that
the increment of phase $\phi _{2}$ for the round trip $OM_{2}O$ in $S$, is 
\begin{equation}
\phi _{2}=k_{AC}^{0}\,l_{0AC}+k_{CA_{2}}^{0}l_{0CA_{2}}=-2\omega t_{M_{2}},
\label{apfi2}
\end{equation}
and $\phi _{2}^{\prime }$ in $S^{\prime }$ is 
\begin{equation}
\phi _{2}^{\prime }=k_{AC}^{0}\,l_{0^{\prime }AC}+k_{CA_{2}}^{0}l_{0^{\prime
}CA_{2}}=-2(\omega /c)(\gamma ct_{M_{2}})=-2\omega (\gamma t_{M_{2}}).
\label{af2cr}
\end{equation}
This is again the result of the traditional analysis, the time
''dilatation,'' $t_{2}^{\prime }=\gamma t_{2}$. For $t_{1}=t_{2}$, i.e., for 
$\overline{L}=L,$ one finally finds the null fringe shift that is obtained
in the traditional analysis $\bigtriangleup N_{e}^{\prime }=\bigtriangleup
N_{e}=0.$ We see that \emph{such a null fringe shift is obtained taking into
account only a part}\ \emph{of the whole phase }$\phi $ \emph{(\ref{phase})
or (\ref{pha2}), and additionally, in that part, }$k_{e}^{0}$\emph{\ is not
changed in transition from} $S$\emph{\ to} $S^{\prime }$. Obviously this
correct result follows from a physically incorrect treatment of the phase $%
\phi $ (\ref{phase}) or (\ref{pha2}). Furthermore it has to be noted that
the usual calculation is always done only in the ''e'' coordinatization.

Since only the part $k_{e}^{0}l_{0e}$\ of the whole phase $\phi $ (\ref
{phase}) or (\ref{pha2}) is taken into account (and also $k_{e}^{0^{\prime
}}=k_{e}^{0}$) the results of the usual ''AT relativity'' calculation are
coordinatization dependent. We explicitly show it using the ''r''
coordinatization.

In the ''r'' coordinatization the increment of phase $\phi _{r}$ is
calculated from $\phi _{r}=k_{r}^{0}g_{00,r}\,l_{r}^{0}$ in $S$ and from $%
\phi _{r}^{\prime }=k_{r}^{0}g_{00,r}\,l_{r}^{0^{\prime }}$ in $S^{\prime }.$
Hence we find that $\phi _{1r}$ for the round trip $OM_{1}O$ in $S$ is 
\begin{equation}
\phi _{1r}=-2(\omega t_{M_{1}}+(2\pi /\lambda )\overline{L}),  \label{f1er}
\end{equation}
and $\phi _{2r}$ for the round trip $OM_{2}O$ in $S$ is 
\begin{equation}
\phi _{2r}=-2(\omega t_{M_{2}}+(2\pi /\lambda )L).  \label{f2er}
\end{equation}
For $\overline{L}=L,$ and consequently $t_{M_{1}}=t_{M_{2}},$ we find that $%
\phi _{1r}-\phi _{2r}=0$, whence $\bigtriangleup N_{r}=0.$ Remark that the
phases $\phi _{1r}$ and $\phi _{2r}$ differ from the corresponding phases $%
\phi _{1e}$ and $\phi _{2e}$ in the ''e'' coordinatization. As shown above
this is not the case when the whole phase $\phi $ (\ref{phase}) or (\ref
{pha2}) is taken into account.

However, in $S^{\prime },$ we find for the same trips that 
\begin{equation}
\phi _{1r}^{\prime }=-2(\gamma \omega t_{M_{1}}(1+\beta )+(2\pi /\lambda )%
\overline{L}),  \label{efcr1}
\end{equation}
\begin{equation}
\phi _{2r}^{\prime }=-2\gamma ^{2}(1+\beta ^{2})(\omega t_{M_{2}}+(2\pi
/\lambda )L).  \label{fcrt2}
\end{equation}
Obviously $\phi _{1r}^{\prime }-\phi _{2r}^{\prime }\neq 0$ and consequently
it leads to \emph{the non-null fringe shift} 
\begin{equation}
\bigtriangleup N_{r}^{\prime }\neq 0,  \label{denrc}
\end{equation}
which holds even in the case when $t_{M_{1}}=t_{M_{2}}.$ This result clearly
shows that the agreement between the usual ''AT relativity'' calculation and
the Michelson-Morley experiment is only an ''apparent'' agreement. It is
achieved by an incorrect procedure and it holds only in the ''e''
coordinatization. We also remark that the traditional analysis, i.e., the
''AT relativity,'' gives different values for the phases, e.g., $\phi _{1e},$
$\phi _{1e}^{\prime },$ $\phi _{1r}$ and $\phi _{1r}^{\prime },$ since only
a part of the whole phase $\phi $ (\ref{phase}) or (\ref{pha2}) is
considered. \emph{These phases are frame and coordinatization dependent
quantities. When the whole phase }$\phi $\emph{\ (\ref{phase}) or (\ref{pha2}%
) is taken into account, i.e., in ''TT relativity,'' all the mentioned
phases are exactly equal quantities; they are the same,} \emph{frame and
coordinatization independent, quantity. \bigskip }

\subsection{The modern laser versions}

The modern laser versions of the Michelson-Morley experiment, e.g., \cite
{jaseja} and \cite{brillet}, are always interpreted according to the ''AT
relativity.'' They rely on highly monochromatic (maser) laser frequency
metrology rather than optical interferometry; the measured quantity is not
the maximum shift in the number of fringes than a beat frequency variation
and the associated (maser) laser-frequency shift. In \cite{jaseja} the
authors recorded the variations in beat frequency between two optical maser
oscillators when rotated through $90^{0}$ in space; the two maser cavities
are placed orthogonally on a rotating table and they can be considered as
two light clocks. It is stated in \cite{jaseja} that the highly
monochromatic frequencies of masers; ''...allow very sensitive detection of
any change in the round-trip optical distance between two reflecting
surfaces.'' and that the comparison of the frequencies of two masers allows:
''...a very precise examination of the isotropy of space with respect to
light propagation.'' The result of this experiment was: ''... there was no
relative variation in the maser frequencies associated with orientation of
the earth in space greater than about 3 kc/sec.'' Similarly \cite{brillet}
compares the frequencies of a He-Ne laser locked to the resonant frequency
of a higly stable Fabry-Perot cavity (the meter-stick, i.e., ''etalon of
length'') and of a $CH_{4}$ stabilized ''telescope-laser'' frequency
reference system. The beat frequency of the isolation laser ($CH_{4}$
stabilized-laser) with the cavity-stabilized laser was the measured
quantity; a beat frequency variation is considered when the direction of the
cavity length is rotated. The authors of \cite{brillet}, in the same way as 
\cite{jaseja}, consider their experiment as: ''isotropy of space
experiment.'' Namely it is stated in \cite{brillet} that: ''Rotation of the
entire electro-optical system maps any cosmic directional anisotropy of
space into a corresponding frequency variation.'' They found a null result,
i.e., a fractional length change of $\bigtriangleup l/l=(1.5\pm 2.5)\times
10^{-15}$ (this is also the fractional frequency shift) in showing the
isotropy of space; this result represented a 4000-fold improvement on the
measurements \cite{jaseja}. In \cite{haugan} the experiment \cite{brillet}
is quoted as the most precise repetition of the Michelson-Morley experiment,
and it is asserted that the experiment \cite{brillet} constrained the two
times, our $t_{1}^{\prime }$ and $t_{2}^{\prime }$, to be equal within a
fractional error of $10^{-15}$. The times $t_{1}^{\prime }$ and $%
t_{2}^{\prime }$ refer to the round-trips in two maser cavities in \cite
{jaseja}, and to the round-trips in the Fabry-Perot cavity in \cite{brillet}%
. These times are calculated in the same way as in the Michelson-Morley
experiment.(see, for example, \cite{haugan}).

The above brief discussion of the experiments \cite{jaseja} and \cite
{brillet}, and the previous analysis of the usual, ''AT relativity,''
calculation of $t_{1}^{\prime }$ and $t_{2}^{\prime }$ in the
Michelson-Morley experiment, suggest that the same remarks as in the
Michelson-Morley experiment hold also for the experiments \cite{jaseja} and 
\cite{brillet}. For example, the reflections of light in maser cavities or
in Fabry-Perot cavity happen on the moving mirrors as in the
Michelson-Morley experiment, which means that the optical paths between the
reflecting ends have to be calculated taking into account the Doppler
effect, i.e., as in Driscoll's procedure \cite{drisc}. In fact, the
interference of the light waves, e.g., the light waves with close
frequencies from two maser cavities in \cite{jaseja}, is always determined
by their phase difference and not only with their frequencies. Also it has
to be noted that the theoretical predictions for the beat frequency
variation are strongly dependent on the chosen synchronization. Hence,
although the measurement of the beat frequency variation is more precise
than the measurement of the shift in the number of fringes, it actually does
not improve the testing of SR. Thus, contrary to the generally accepted
opinion, the experiments \cite{jaseja} and \cite{brillet} do not confirm the
validity of the usual ''AT relativity.''

Regarding the ''TT relativity,'' the modern laser versions \cite{jaseja} and 
\cite{brillet} of the Michelson-Morley experiment are incomplete experiments
(only the beat frequency variation is measured) and cannot be compared with
the theory; in the ''TT relativity'' the same 4D quantity has to be
considered in relatively moving IFRs and \emph{the frequency, taken alone,
is not a 4D quantity}.

\section{THE KENNEDY-THORNDIKE TYPE EXPERIMENTS}

In the Kennedy-Thorndike experiment \cite{kenne} a Michelson interferometer
with unequal armlengths was employed and they looked for possible diurnal
and annual variations in the difference of the optical paths due to the
motion of the interferometer with respect to the preferred frame. The
measured quantity was, as in the Michelson-Morley experiment, the shift in
the number of fringes, and in \cite{kenne} the authors also found that was
no observable fringe shift. We shall not discuss this experiment since the
whole consideration is completely the same as in the case of the
Michelson-Morley experiment, and, consequently, the same conclusion holds
also here, i.e., the experiment \cite{kenne} does not agree with the ''AT
relativity,'' but directly proves the ''TT relativity.'' A modern version of
the Kennedy-Thorndike experiment was carried out in \cite{hils}, and the
authors stated: ''We have performed the physically equivalent measurement
(with the Kennedy-Thorndike experiment, my remark) by searching for a
sidereal 24-h variation in the frequency of a stabilized laser compared with
the frequency of a laser locked to a stable cavity.'' The result was: ''No
variations were found at the level of $2\times 10^{-13}."$ Also they
declared: ''This represents a 300-fold improvement over the original
Kennedy-Thorndike experiment and \emph{allows the Lorentz transformations to
be deduced entirely from experiment at an accuracy level of 70 ppm.'' }(my
emphasis) The experiment \cite{hils} is of the same type as the experiment 
\cite{brillet}, and neither the experiment \cite{brillet} is physically
equivalent to the Michelson-Morley experiment, as shown above, nor, contrary
to the opinion of the authors of \cite{hils}, the experiment \cite{hils} is
physically equivalent to the Kennedy-Thorndike experiment; the measurement
of the beat frequency variation is not equivalent to the measurement of the
change in the phase difference (in terms of the measurement of the shift in
the number of fringes). Namely such equivalence can exist only in the usual
''AT relativity'' treatment since there the phase difference is determined
only by the time difference. And, additionally, the Michelson-Morley and the
Kennedy-Thorndike experiments can be compared both with the ''AT
relativity'' and the ''TT relativity'', while the modern laser versions \cite
{brillet}, \cite{jaseja} and \cite{hils} of these experiments are incomplete
experiments from the ''TT relativity'' viewpoint and cannot be compared with
the ''TT relativity.'' Furthermore, the ''TT relativity'' deals with the
covariant 4D Lorentz transformations $L^{a}{}_{b}$ (\ref{fah}), or with
their representations $L^{\mu ^{\prime }}{}_{\nu ,e}$ (\ref{lorus}) in the
''e'' coordinatization and with $L^{\mu ^{\prime }}\,_{\nu ,r}$ (\ref{elr})
in the ''r'' coordinatization, and none of them can be deduced from the
experiment \cite{hils}. Thus the treatment of the Michelson-Morley
experiment with true tensor quantities from \cite{tom2} and Sec. 5.1 here
reveals that \emph{the relevant quantity for the measurements both in the
Michelson-Morley and the Kennedy-Thorndike type experiments is the phase (%
\ref{phase}) and in the experiments it has to be determined according to the
relation (\ref{pha2}).}

\section{THE IVES-STILLWEL TYPE EXPERIMENTS}

Ives and Stilwell \cite{ives} performed a precision Doppler effect
experiment in which they used a beam of excited hydrogen molecules as a
moving light source. The frequencies of the light emitted parallel and
antiparallel to the beam direction were measured by a spectograph (at rest
in the laboratory). The measured quantity in this experiment is 
\begin{equation}
\bigtriangleup f/f_{0}=(\bigtriangleup f_{b}-\bigtriangleup f_{r})/f_{0},
\label{frdop}
\end{equation}
where $f_{0}$ is the frequency of the light emitted from resting atoms. $%
\bigtriangleup f_{b}=\left| f_{b}-f_{0}\right| $ and $\bigtriangleup
f_{r}=\left| f_{r}-f_{0}\right| ,$ where $f_{b}$ is the blue-Doppler-shifted
frequency that is emitted in a direction parallel to $\mathbf{v}$ ($\mathbf{v%
}$ is the velocity of the atoms relative to the laboratory), and $f_{r}$ is
the red-Doppler-shifted frequency that is emitted in a direction opposite to 
$\mathbf{v.}$ The quantity $\bigtriangleup f/$ $f_{0}$ measures the extent
to which the frequency of the light from resting atoms fails to lie halfway
between the frequencies $f_{r}$ and $f_{b}.$ In terms of wavelengths the
relation (\ref{frdop}) can be written as 
\begin{equation}
\bigtriangleup \lambda /\lambda _{0}=(\bigtriangleup \lambda
_{r}-\bigtriangleup \lambda _{b})/\lambda _{0},  \label{ladop}
\end{equation}
where $\bigtriangleup \lambda _{r}=\left| \lambda _{r}-\lambda _{0}\right| $
and $\bigtriangleup \lambda _{b}=\left| \lambda _{b}-\lambda _{0}\right| ,$
and, as we said, $\lambda _{r}$ and $\lambda _{b}$ are the wavelengths
shifted due to the Doppler effect to the ''red'' and ''blue'' regions of the
spectrum. In that way Ives and Stilwell replaced the difficult problem of
the precise determination of the wavelength with much simpler problem of the
determination of the asymmetry of shifts of the ''red'' and ''blue'' shifted
lines with respect to the unshifted line. They \cite{ives} showed that the
measured results agree with the formula predicted by the traditional
formulation of SR, i.e., the usual ''AT relativity,'' and not with the
classical nonrelativistic expression for the Doppler effect. Let us explain
it in more detail.

\subsection{The ''AT relativity'' calculation}

In the ''AT relativity'' one usually starts with the Lorentz transformation
of the basis components $k^{\mu }(\omega /c,\mathbf{k=n}\omega /c\mathbf{)}$
of the 4-vector $k^{a}$ of the light wave from an IFR $S$ to the relatively
moving (along the common $x,x^{\prime }-$axes) IFR $S^{\prime }$. Note that
only the ''e'' coordinatization is used in such traditional treatment. Then
the Lorentz transformation in the ''e'' coordinatization of $k^{\mu }$ can
be written as 
\begin{equation}
k^{0^{\prime }}=\omega ^{\prime }/c=\gamma (\omega /c-\beta
k^{1}),k^{1^{\prime }}=\gamma (k^{1}-\beta \omega /c),k^{2^{\prime
}}=k^{2},k^{3^{\prime }}=k^{3},  \label{lordop}
\end{equation}
or in terms of the unit wave vector $\mathbf{n}$ (which is in the direction
of propagation of the wave) 
\begin{equation}
\omega ^{\prime }=\gamma \omega (1-\beta n^{1}),n^{1^{\prime
}}=N(n^{1}-\beta ),n^{2^{\prime }}=(N/\gamma )n^{2},n^{3^{\prime
}}=(N/\gamma )n^{3},  \label{odop}
\end{equation}
where $N=(1-\beta n^{1})^{-1}.$ Now comes the main point in the derivation.
Although the Lorentz transformation of the basis components $k^{\mu }$ of
the 4-vector $k^{a}$ from $S$ to $S^{\prime },$ Eqs.(\ref{lordop}) and (\ref
{odop}), transforms all four components of $k^{\mu }$ the usual ''AT
relativity'' treatment considers the transformation of the temporal part of $%
k^{\mu },$ i.e., the frequency, as independent of the transformation of the
spatial part of $k^{\mu },$ i.e., the unit wave vector $\mathbf{n.}$ Thus
the ''AT relativity'' deals with two \emph{independent }physical phenomena -
the Doppler effect and the aberration of light. (Recall that we have already
met such omission of one part of the Lorentz transformation of a 4-vector
(written in the ''e'' coordinatization) in the derivation of the expressions
for the Lorentz contraction (\ref{apcon}) and the dilatation of time (\ref
{tidil}) in Sec. 2.2.) We note once again that such distinction is possible
only in the ''e'' coordinatization; in the ''r'' coordinatization the metric
tensor $g_{\mu \nu ,r}$ is not diagonal and consequently the separation of
the temporal and spatial parts does not exist. Thus the ''AT relativity''
calculation is restricted to the ''e'' coordinatization. In agreement with
such theoretical treatment the existing experiments (including the modern
experiments based on collinear laser spectroscopy; see, e.g., \cite
{mcgow,riis,klein}, or the review \cite{kretz}) are designed in such a way
to measure either the Doppler effect or the aberration of light. Let us
write the above transformation in the form from which one can determine the
quantities in (\ref{ladop}) and then compare them with the experiments. The
spectograph is at rest in the laboratory (the $S$ frame) and the light
source (at rest in the $S^{\prime }$ frame) is moving with $\mathbf{v}$
relative to $S.$ Then in the usual ''AT relativity'' approach \emph{only the
first relation from (\ref{lordop}), or (\ref{odop}), is used,} which means
that, in the same way as shown in previous cases, \emph{the ''AT
relativity'' deals with two different quantities in 4D spacetime, here }$%
\omega $\emph{\ and }$\omega ^{\prime }$\emph{.} Then writting the
transformation of the temporal part of $k^{\mu },$ i.e., of $\omega ,$ in
terms of the wavelength $\lambda $ we find 
\begin{equation}
\lambda =\gamma \lambda _{0}(1-\beta \cos \theta ),  \label{lam1}
\end{equation}
where $\lambda $ is the wavelength received in the laboratory from the
moving source (the shifted line), $\lambda _{0}$ ($=\lambda ^{\prime }$) is
the natural wavelength (the unshifted line) and $\theta $ is the angle of $%
\mathbf{k}$ relative to the direction of $\mathbf{v}$ as measured in the
laboratory. The nonrelativistic treatment of the Doppler effect predicts $%
\lambda =\lambda _{0}(1-\beta \cos \theta ),$ and in the classical case the
Doppler shift does not exist for $\theta =\pi /2$. This transverse Doppler
effect ($\theta =\pi /2,$ $\lambda =\gamma \lambda _{0},$ or $\nu =\nu
_{0}/\gamma $) is always, in the traditional, ''AT relativity,'' approach
considered to be a direct consequence of the time dilatation; it is asserted
(e.g. \cite{Kittel}) that the frequencies must be related as the inverse of
the times in the usual relation for the time dilatation $\bigtriangleup
t=\bigtriangleup t_{0}\gamma $. It is usually interpreted \cite{kretz}:
''The Doppler shift experiments ... compare the rates of two ''clocks'' that
are in motion relative to each other. \emph{They measure time dilatation}
(my emphasis) and can test the validity of the special relativity in this
respect.'' Similarly it is declared in \cite{mcgow}: ''The experiment
represents a more than tenfold improvement over other Doppler shift
measurements and \emph{verifies the time dilation effect} (my emphasis) at
an accuracy level of 2.3 ppm.'' Obviously, as we said, the Doppler shift
experiments are theoretically analysed only by means of the ''AT
relativity,'' which treats the transformation of the temporal part of $%
k^{\mu }$ as independent of the transformation of the spatial part of $%
k^{\mu },$ and moreover completely neglects the Lorentz transformation of
the spatial part of $k^{\mu }.$

In the Ives and Stilwell type experiments the measurements are conducted at
symmetric observation angles $\theta $ and $\theta +180^{0};$ particularly
in \cite{ives} $\theta $ is chosen to be $\simeq 0^{0}$. The wavelength in
the direction of motion is obtained from (\ref{lam1}) as $\lambda
_{b}=\gamma \lambda _{0}(1-\beta \cos \theta ),$ while that one in the
opposite direction (the angle $\theta +180^{0}$) is $\lambda _{r}=\gamma
\lambda _{0}(1+\beta \cos \theta ),$ and then $\bigtriangleup \lambda
_{b}=\left| \lambda _{b}-\lambda _{0}\right| =\left| \lambda _{0}(1-\gamma
+\beta \gamma \cos \theta )\right| ,$ $\bigtriangleup \lambda _{r}=\left|
\lambda _{r}-\lambda _{0}\right| =\left| \lambda _{0}(\gamma -1+\beta \gamma
\cos \theta )\right| ,$ and the difference in shifts is 
\begin{equation}
\bigtriangleup \lambda =\bigtriangleup \lambda _{r}-\bigtriangleup \lambda
_{b}=2\lambda _{0}(\gamma -1)\simeq \lambda _{0}\beta ^{2},  \label{dellam}
\end{equation}
where the last relation holds for $\beta \ll 1.$ Note that the redshift due
to the transverse Doppler effect ($\lambda _{0}\beta ^{2}$) is independent
on the observation angle $\theta $. In the nonrelativistic case $%
\bigtriangleup \lambda =0$, the transverse Doppler shift is zero. Ives and
Stilwell found the agreement of the experimental results with the relation (%
\ref{dellam}) and not with the classical result $\bigtriangleup \lambda =0.$

However, a more careful analysis shows that the agreement between the ''AT
relativity'' prediction Eq.(\ref{dellam}) and the experiments \cite{ives}
is, contrary to the general belief, only an ''apparent'' agreement and not
the ''true'' one. This agreement actually happens for the following reasons.
First, the theoretical result (\ref{dellam}) is obtained in the ''e''
coordinatization in which one can speak about the frequency $\omega $ and
the wave vector $\mathbf{k}$ as well-defined quantities. Using the matrix $%
T^{\mu }{}_{\nu ,r}$ (\ref{teer}) which transforms the ''e''
coordinatization to the ''r'' coordinatization, $k_{r}^{\mu }=T^{\mu
}{}_{\nu ,r}k_{e}^{\nu }$ (only the components are considered), one finds $%
k_{r}^{0}=k_{e}^{0}-k_{e}^{1}-k_{e}^{2}-k_{e}^{3},\quad k_{r}^{i}=k_{e}^{i},$
whence we conclude that in the ''r'' coordinatization the theoretical
predictions for \emph{the components} of a 4-vector, i.e., for $\lambda ,$
will be quite different but in the ''e'' coordinatization, i.e., but the
result (\ref{dellam}), and thus not in the agreemement with the experiment 
\cite{ives}. Further, \emph{the specific choice of} $\theta $ ($\theta
\simeq 0^{0})$\emph{\ in the experiments \cite{ives} is the next reason for
the agreement with the ''AT relativity'' result (\ref{dellam}).} Namely, 
\emph{if} $\theta =0^{0}$ \emph{then} $n^{1}=1,$ $n^{2}=n^{3}=0$, and $%
k^{\mu }$ is $(\omega /c,\omega /c,0,0\mathbf{).}$ From (\ref{lordop}) or (%
\ref{odop}) one finds that \emph{in }$S^{\prime }$ \emph{too} $\theta
^{\prime }=0^{0},$ $n^{1^{\prime }}=1$ and $n^{2^{\prime }}=n^{3^{\prime }}=0
$ (the same holds for $\theta =180^{0},$ $n^{1}=-1,$ $n^{2}=n^{3}=0$, then $%
\theta ^{\prime }=180^{0}$ and $n^{1^{\prime }}=-1,$ $n^{2^{\prime
}}=n^{3^{\prime }}=0$). In the experiments \cite{ives} the emitter is the
moving ion (its rest frame is $S^{\prime }$), while the observer is the
spectrometer at rest in the laboratory (the $S$ frame). Since in \cite{ives}
the angle of the ray emitted by the ion at rest is chosen to be $\theta
^{\prime }=0^{0}$ ($180^{0}$), then the angle of this ray measured in the
laboratory, where the ion is moving, will be the same $\theta =0^{0}$ ($%
180^{0}$). (Similarly happens in the modern versions \cite{mcgow,klein} of
the Ives-Stilwell experiment; the experiments \cite{mcgow,klein} make use of
an atomic or ionic beam as a moving light analyzer (the accelerated ion is
the ''observer'') and two collinear laser beams (parallel and antiparallel
to the particle beam) as light sources (the emitter), which are at rest in
the laboratory.) From this consideration we conclude that in these
experiments one can consider only the Doppler effect, that is, the
transformation of $\omega $ (the temporal part of $k^{\mu };$ the component
form of the true 4-vector $k^{a}$ in the ''e'' coordinatization), and not
the aberration of light, i.e., the transformation of $\mathbf{n,}$ i.e., $%
\mathbf{k,}$ (the spatial part of $k^{\mu }$). Because of that they found
the agreement between the relation (\ref{lam1}) (or (\ref{dellam})) with the
experiments. However, the relations (\ref{lordop}) and (\ref{odop}) reveal
that in the case of an arbitrary $\theta $ the transformation of the
temporal part of $k^{\mu }$ cannot be considered as independent of the
transformation of the spatial part. This means that in such case one cannot
expect that the relation (\ref{dellam}), taken alone, will be in agreement
with the experiments performed at some arbitrary $\theta .$ Such experiments
were, in fact, recently conducted and we discuss them here.

Pobedonostsev and collaborators \cite{pobed} performed the Ives-Stilwell
type experiment but improved the experimental setup and, what is
particularly important, the measurements were conducted at symmetric
observation angles $77^{0}$ and $257^{0},$ which are different from $0^{0}$
(and $180^{0}$). The measurement was done with a beam of $H_{2}^{+}$ ions at
energies $175,180,210,225,260$ and $275$ $keV.$ The radiation from hydrogen
atoms in excited state, which are formed as a result of disintegration of
accelerated $H_{2}^{+},$ was observed. The radiation from the moving
hydrogen atoms, giving the Doppler shifted lines, was observed together with
the radiation from the resting atoms existing in the same working volume,
and giving an unshifted line. The similar work was reported in \cite{pobe2}
in which a beam of $H_{3}^{+}$ ions at energy $310$ $keV$ was used and the
measurements were conducted at symmetric observation angles $82^{0}$ and $%
262^{0}.$ The results of the experiments \cite{pobed} and \cite{pobe2}
markedly differed from all previous experiments that were performed at
observation angles $\theta =0^{0}$ (and $180^{0}$). Therefore in \cite{pobe2}
Pobedonostsev declared: \emph{''In comparing the wavelength of Doppler
shifted line from a moving emitter with the wavelength of an identical
static emitter, the experimental data corroborate the classical formula for
the Doppler effect, not the relativistic one.''} Thus, instead of to find
the ''relativistic'' result $\bigtriangleup \lambda \simeq \lambda _{0}\beta
^{2}$ (\ref{dellam}), (actually the ''AT relativity'' result), they found
the classical result $\bigtriangleup \lambda \simeq 0,$ i.e., they found
that the redshift due to the transverse Doppler effect ($\lambda _{0}\beta
^{2}$) \emph{is dependent} on the observation angle $\theta $. This
experimental result strongly support our assertion that the agreement
between the ''AT relativity'' and the Ives-Stilwell type experiments is only
an ''apparent'' agreement and not the ''true'' one.

\subsection{The ''TT relativity'' approach}

As already said in the ''TT relativity'' neither the Doppler effect nor the
aberration of light exist separately as well defined physical phenomena. As
shown in \cite{tom1,tom2} and Sec. 2.2 here (see (\ref{tidil}) and the
discussion there) in the 4D spacetime the temporal distances (e.g., $\tau
_{E}$ and $\tau _{\mu }$ from Sec. 4.2) refer to different quantities, which
are not connected by the Lorentz transformation. The same happens with $%
\omega $ and $\omega ^{\prime }$ as the temporal parts of $k^{\mu },$ the
component form of $k^{a}$ in the ''e'' coordinatization . And, as Gamba \cite
{gamba} stated, the fact that the measurements of such quantities were made
by \emph{two} observers does not mean that relativity has something to do
with the problem. In the ''TT relativity'' the entire 4D quantity, the true
tensor or the CBGQ, has to be considered both in the theory and \emph{in
experiments}. Therefore, in order to theoretically discuss the experiments
of the Ives-Stilwell type we choose as the relevant quantity the wave vector 
$k^{a},$ the geometric quantity, which can be written in the
coordinate-based geometric language as the relation (\ref{kad}), $%
k^{a}=k^{\mu ^{\prime }}e_{\mu ^{\prime }}=k^{\mu }e_{\mu }=k_{r}^{\mu
^{\prime }}r_{\mu ^{\prime }}=k_{r}^{\mu }r_{\mu }.$ Equivalently one can
consider its square for which it holds that 
\begin{equation}
k^{a}g_{ab}k^{b}=0;  \label{kadop}
\end{equation}
this expression is a Lorentz scalar and it is also independent of the choice
of the coordinatization. The relations (\ref{kad}) and (\ref{kadop}) show
that we can calculate $k^{a}$ (or $k^{a}g_{ab}k^{b}$) in the ''e''
coordinatization and in the rest frame of the emitter (the $S^{\prime }$
frame); the emitter is the ion moving in $S,$ the rest frame of the
spectrometer, i.e., in the laboratory frame. In other permissible
coordinatizations and in other relatively moving IFRs these quantities will
be exactly the same as in $S^{\prime }$ and the ''e'' coordinatization. That
is a great practical advantage of the true tensor formulation of SR; \emph{%
when the whole (including the basis) 4D tensor quantity is considered then
it is an invariant quantity. }

First we consider the experiments \cite{pobed} and \cite{pobe2} since they
showed the disagreement with the traditional theory, i.e., with the ''AT
relativity.'' Then $k^{a}$ in the ''e'' coordinatization and in $S^{\prime }$
is represented by the CBGQ $k^{\mu ^{\prime }}e_{\mu ^{\prime }}$ whence the
components $k^{\mu ^{\prime }}$ are $k^{\mu ^{\prime }}=(\omega ^{\prime
}/c)(1,\cos \theta ^{\prime },\sin \theta ^{\prime },0\mathbf{)}$ and $%
k^{\mu ^{\prime }}k_{\mu ^{\prime }}=0.$ The observer (the spectrometer) in
the laboratory frame will look at \emph{the same 4D quantity} $k^{a},$ or
equivalently the CBGQ $k^{\mu }e_{\mu }$, and find $k^{\mu },$ the Lorentz
transformed component form in the ''e'' coordinatization of the wave vector $%
k^{\mu }e_{\mu },$ as 
\[
k^{\mu }=\left[ \gamma (\omega ^{\prime }/c)(1+\beta \cos \theta ^{\prime
}),\gamma (\omega ^{\prime }/c)(\cos \theta ^{\prime }+\beta ),(\omega
^{\prime }/c)\sin \theta ^{\prime },0\right] ,
\]
whence $k^{\mu }k_{\mu }$ is also $=0.$ From that transformation one can
find that 
\[
n^{1}=(n^{1^{\prime }}+\beta )/(1+\beta n^{1^{\prime }}),n^{2}=n^{2^{\prime
}}/\gamma (1+\beta n^{1^{\prime }}),n^{3}=n^{3^{\prime }}/\gamma (1+\beta
n^{1^{\prime }}),
\]
or that 
\[
\sin \theta =\sin \theta ^{\prime }/\gamma (1+\beta \cos \theta ^{\prime
}),\cos \theta =(\cos \theta ^{\prime }+\beta )/(1+\beta \cos \theta
^{\prime }),
\]
\begin{equation}
\tan \theta =\sin \theta ^{\prime }/\gamma (\beta +\cos \theta ^{\prime }).
\label{tanab}
\end{equation}
The relations (\ref{tanab}) reveal that not only $\omega $ is changed (the
Doppler effect) when going from $S^{\prime }$ to $S$ but also the angle of $%
\mathbf{k}$ relative to the direction of $\mathbf{v}$ is changed (the
aberration of light). This means that if the observation of the unshifted
line (i.e., of the frequency $\omega ^{\prime }=\omega _{0}$ from the atom
at rest) is performed at an observation angle $\theta ^{\prime }$ in $%
S^{\prime },$ the rest frame of the emitter, then \emph{the same light wave}
(from the same but now moving atom) will have the shifted frequency $\omega $
and \emph{will be seen }at an observation angle $\theta $ (generally, $\neq
\theta ^{\prime }$) in $S,$ the rest frame of the spectrometer. In $%
S^{\prime }$ the quantities $\omega ^{\prime }$ and $\theta ^{\prime }$
define the CBGQ $k^{\mu ^{\prime }}e_{\mu ^{\prime }},$ and this propagation
4-vector satisfies the relation $k^{\mu ^{\prime }}k_{\mu ^{\prime }}=0,$
which is the representation of the relation (\ref{kadop}) in the ''e''
coordinatization and in the $S^{\prime }$ frame. The quantities $\omega
^{\prime }$ and $\theta ^{\prime }$ (that define the corresponding $k^{\mu
^{\prime }}e_{\mu ^{\prime }}$ in $S^{\prime }$) are connected with the
corresponding $\omega $ and $\theta $ (that define the corresponding $k^{\mu
}e_{\mu }$ in $S$) by means of the Lorentz transformation $L^{\mu ^{\prime
}}{}_{\nu ,e}$ (\ref{lorus}) (and its inverse) of $k^{\mu ^{\prime }}e_{\mu
^{\prime }}.$ Then $k^{\mu }e_{\mu }$ is such that it also satisfies the
relation $k^{\mu }k_{\mu }=0,$ the representation of (\ref{kadop}) in the
''e'' coordinatization and now in the $S$ frame. The authors of the
experiments \cite{pobed} (and \cite{pobe2}) \emph{made the observation of
the radiation from the atom at rest (the unshifted line) and from a moving
atom at the same observation angle.} The preceding discussion shows that if
they succeeded to see $\omega ^{\prime }=\omega _{0}$ (i.e., $\lambda _{0}$)
from the atom at rest at some symmetric observation angles $\theta ^{\prime }
$ ($\neq 0$) and $\theta ^{\prime }+180^{0}$ (i.e., some $k^{\mu ^{\prime
}}e_{\mu ^{\prime }}$) then they could not see the assymetric Doppler shift
(from moving atoms) at the same angles $\theta =\theta ^{\prime }$ (and $%
\theta +180^{0}=\theta ^{\prime }+180^{0}$). The Lorentz transformation does
not connect such quantities. This was the reason that they detected $%
\bigtriangleup \lambda \simeq 0$ and not $\bigtriangleup \lambda \simeq
\lambda _{0}\beta ^{2}.$ But we expect that the result $\bigtriangleup
\lambda \simeq \lambda _{0}\beta ^{2}$ can be seen \emph{if the similar
measurements of the frequencies, i.e., the wavelengths, of the radiation
from moving atoms would be performed not at }$\theta =\theta ^{\prime }$ 
\emph{but at} $\theta $ \emph{determined by the relation (\ref{tanab}). Only
in that case one will make measurement of the same quantity} $k^{a}=k^{\mu
^{\prime }}e_{\mu ^{\prime }}=k^{\mu }e_{\mu }$ \emph{from two different
relatively moving IFRs.}

Recently, Bekljamishev \cite{bekla} came to the same conclusions (but
dealing only with the component form in the ''e'' coordinatization) and
explained the results of the experiments \cite{pobed} and \cite{pobe2}
taking into account the aberration of light together with the Doppler
effect. It is argued in \cite{bekla} that Eq.(\ref{lam1}) for the Doppler
effect can be realized only when the condition for the aberration angle is
fulfilled, 
\begin{equation}
\bigtriangleup \theta =\beta \sin \theta ^{\prime },  \label{aber}
\end{equation}
where $\bigtriangleup \theta =\theta ^{\prime }-\theta ,$ and $\beta $ is
taken to be $\beta \ll 1.$ The relation (\ref{aber}) directly follows from
the expression for $\sin \theta $ in (\ref{tanab}) taking that $\beta \ll 1.$
\emph{The assymetric shift will be seen when the collimator assembly is
tilted at a velocity dependent angle }$\bigtriangleup \theta .$ Instead of
to work, as usual, with the arms of the collimator at fixed angles $\theta $
and $\theta +180^{0},$ Bekljamishev \cite{bekla} proposed that the
collimator assembly must be constructed in such a way that there is the
possibility of the correction of the observation angles independently for
both arms; for example, the arm at angle $\theta $ ($\theta +180^{0}$) has
to be tilted clockwise (counter-clockwise) by the aberration angle $%
\bigtriangleup \theta .$ Otherwise the assymetry in the Doppler shifts will
not be observed. Thus the experiments \cite{pobed} and \cite{pobe2} would
need to be repeated taking into account Bekljamishev's proposition. \emph{%
The positive result for the Doppler shift }$\bigtriangleup \lambda $\emph{\ (%
\ref{dellam}), when the condition for the aberration angle }$\bigtriangleup
\theta $\emph{\ (\ref{aber}) is fulfilled, will definitely show that it is
not possible to treat the Doppler effect and the aberration of light as
separate, well-defined, effects, i.e., that it is the ''TT relativity,'' and
not the ''AT relativity,'' which correctly explains the experiments that
test SR.}

\section{CONCLUSIONS\ AND\ DISCUSSION}

In the first part of this paper we have discussed and exposed the main
differences between three theoretical formulations of SR, the ''TT
relativity,'' the covariant approach to SR and the ''AT relativity.'' In the
second part we have presented the comparison of these formulations with the
experiments. The analysis of the experiments which test SR shows that they
agree with the predictions of the ''TT relativity'' and not, as usually
supposed, with those of the ''AT relativity.''

In the ''muon'' experiment the fluxes of muons on a mountain, $N_{m}$, and
at sea level, $N_{s}$, are measured. The ''AT relativity'' predicts
different values of the flux $N_{s}$ (for the same measured $N_{m}$) in
different synchronizations, but the measured $N_{s}$ is of course
independent of the chosen coordinatization. Further, for some
synchronizations these predicted values of the flux at sea level $N_{s}$ are
quite different than the measured ones. The reason for such disagreement, as
explained in the theoretical part of this paper, Secs. 2, 2.1 and 2.2, is
that in the usual, ''AT relativity,'' analysis of the ''muon'' experiment,
for example, the lifetimes $\tau _{E}$ and $\tau _{\mu }$ are considered to
refer to the same temporal distance (the same quantity) measured by the
observers in two relatively moving IFRs. But the transformation connecting $%
\tau _{E}$ and $\tau _{\mu }$ (the dilatation of time (\ref{tidil})) is only 
\emph{a part} of the Lorentz transformation written in the ''e''
coordinatization, and, actually, $\tau _{E}$ and $\tau _{\mu }$ refer to
different quantities in 4D spacetime. Although their measurements were made
by\emph{\ two} observers, the relativity has nothing to do with the problem,
since $\tau _{E}$ and $\tau _{\mu }$ are different 4D quantities. \emph{The
''TT relativity,'' in contrast to the ''AT relativity,'' completely agrees
with the ''muon'' experiments in all IFRs and all permissible
coordinatizations. }In the ''TT relativity''\emph{\ the same 4D quantity} (a
true tensor or a CBGQ) is considered in different IFRs and different
coordinatizations; instead of to work with $\tau _{E}$ and $\tau _{\mu }$
the ''TT relativity'' deals with the spacetime length $l$ and the distance
4-vector $l_{AB}^{a}$ and formulate the radioactive-decay law in terms of
invariant quantities, i.e., the true tensors or the CBGQs, Eqs. (\ref{radTT}%
), (\ref{radlaw}) and (\ref{radla1}).

In the Michelson-Morley experiment the traditional, ''AT relativity,''
derivation of the fring shift $\bigtriangleup N$ deals only with the
calculation, in the ''e'' coordinatization, of $t_{1}$ and $t_{2}$ (in $S$
and $S^{\prime }$)$,$ which are the times required for the complete trips $%
OM_{1}O$ and $OM_{2}O$ along the arms of the Michelson-Morley
interferometer. The null fringe shift obtained with such calculation is only
in an ''apparent,'' not ''true,'' agreement with the observed null fringe
shift, since this agreement was obtained by an incorrect procedure. Namely
it is supposed in such derivation that, e.g., $t_{1}$ and $t_{1}^{\prime }$
refer to the same quantity measured by the observers in relatively moving
IFRs $S$ and $S^{\prime }$ that are connected by the Lorentz transformation.
However the relation $t_{1}^{\prime }=\gamma t_{1},$ as shown in Secs. 2,
2.1 and 2.2, is not the Lorentz transformation of some 4D quantity, and $%
t_{1}^{\prime }$\ and $t_{1}$\ do not correspond to the same 4D quantity
considered in $S^{\prime }$ and $S$ respectively. Our ''TT relativity,'' in
contrast to the ''AT relativity'' calculations, deals always with the true
tensor quantities or the CBGQs; in the Michelson-Morley experiment it is the
phase (\ref{phase}) $\phi =k^{a}g_{ab}l^{b}$ defined as the true tensor
quantity, or equivalently the phase (\ref{pha2}) defined as the CBGQ. \emph{%
The ''TT relativity'' calculations yields the observed null fringe shift (%
\ref{delshif}) and that result holds for all IFRs and all coordinatizations.}
In addition we have shown that the usual ''AT relativity'' actually deals
only with the part $k^{0}l_{0}$ of the whole phase $\phi ,$ (\ref{phase}) or
(\ref{pha2}). This contribution $k^{0}l_{0}$ is considered in the
interferometer rest frame $S,$ while in the $S^{\prime }$ frame, in which
the interferometer is moving, the usual ''AT relativity'' takes into account
only the contribution $k^{0}l_{0^{\prime }}$; the $k^{0}$ factor is taken to
be the same in $S$ and $S^{\prime }$ frames (all is done only in the
''e''coordinatization). Thus in the usual ''AT relativity'' two different
quantities $k_{e}^{0}l_{0e}$ and $k_{e}^{0}l_{0^{\prime }e}$ (only the parts
of the phase (\ref{phase}) or (\ref{pha2})) are considered to be the same 4D
quantity for observers in $S$ and $S^{\prime }$ frames, and these quantities
are considered to be connected by the Lorentz transformation. Such an
incorrect procedure then caused an apparent (not true) agreement of the
traditional analysis with the results of the Michelson-Morley experiment.
Since only a part of the whole phase $\phi $ (\ref{phase}) or (\ref{pha2})
is considered the traditional result is synchronization, i.e.,
coordinatization, dependent results. The agreement between the traditional
analysis and the experiment exists only when Einstein's synchronization of
distant clocks is used and not for another synchronization. This is also
proved in Sec. 4.1, where the non-null fringe shift (\ref{denrc}) is found
for the ''r'' coordinatization. The improved ''AT relativity'' calculation
of the fringe shift from \cite{drisc} (again in the ''e'' coordinatization)
takes into account the changes in frequencies due to the Doppler effect and
finds a ''surprising'' non-null fringe shift (\ref{shift}). We have shown in
Sec. 4.1 that the non-null theoretical result for the fringe shift (\ref
{shift}) from \cite{drisc} is easily obtained from our ''TT relativity''
approach taking only the product $k_{e}^{0^{\prime }}l_{0^{\prime }e}$\ in
the calculation of the increment of phase $\phi _{e}^{\prime }$\ in $%
S^{\prime }$ in which the apparatus is moving. Thus again as in the usual
''AT relativity'' calculation two different quantities $k_{e}^{0}l_{0e}$ and 
$k_{e}^{0^{\prime }}l_{0^{\prime }e}$ (only the parts of the phase (\ref
{phase}) or (\ref{pha2})) are considered to be the same 4D quantity for
observers in $S$ and $S^{\prime }$ frames, and consequently that these two
quantities are connected by the Lorentz transformation. Since only a part $%
k_{e}^{0^{\prime }}l_{0^{\prime }e}$ of the whole 4D tensor quantity $\phi $
(\ref{phase}) or (\ref{pha2}) is considered the non-null fringe shift (\ref
{shift}) can be shown to be quite different in another coordinatization,
e.g., in the ''r'' coordinatization (see \cite{tom2}).

The same conclusions can be drawn for the Kennedy-Thorndike type experiments.

In the Ives-Stilwell type experiments the agreement between the ''AT
relativity'' calculation for the Doppler effect and the experiments is again
only an ''apparent'' agreement and not the ''true'' one. Namely the
transverse Doppler shift ($\lambda _{0}\beta ^{2}$, (\ref{dellam})) is
obtained in the ''e'' coordinatization in which one can speak about the
frequency $\omega $ and the wave vector $\mathbf{k}$ as well-defined
quantities. Further in the usual ''AT relativity'' approach only the
transformation of $\omega $ (the temporal part of $k^{\mu }$) is considered,
while the aberration of light, i.e., the transformation of $\mathbf{n,}$
i.e., $\mathbf{k,}$ (the spatial part of $k^{\mu }$) is neglected. ($k^{\mu }
$ is the component form in the ''e'' coordinatization of the true tensor $%
k^{a}$ (\ref{kad}).) Thus in this case too the ''AT relativity'' deals with
two different quantities in 4D spacetime, $\omega $\ and $\omega ^{\prime }$%
, which are not connected by the Lorentz transformation. However, for the
specific choice of the observation angles $\theta ^{\prime }=0^{0}$ ($180^{0}
$) in $S^{\prime }$ (the rest frame of the emitter), one finds from the
transformation of $k^{\mu }$ that $\theta $ in $S$ is again $=0^{0}$ ($%
180^{0}$). Since in the experiments \cite{ives}, and its modern versions 
\cite{mcgow,klein}, just such angles were chosen, it was possible to
consider only the transformation of $\omega $, i.e., only the Doppler
effect, and not the concomitant aberration of light. Because of that they
found the agreement between the relation (\ref{lam1}) (or (\ref{dellam}))
and the experiments. When the experiments were performed at observation
angles $\theta \neq 0^{0}$ (and $180^{0}$), as in \cite{pobed} and \cite
{pobe2}, the results disagreed with the ''AT relativity'' calculation which
takes into account only the transformation of $\omega $, i.e., only the
Doppler effect. Furthermore, since the ''AT relativity'' calculation deals
only with a part of the whole 4D quantity $k^{a}$ (\ref{kad}), the agreement
with the experiments will not exist in, e.g., the ''r'' coordinatization.
The ''TT relativity'' calculation considers the whole 4D quantity, the wave
vector $k^{a}$ (\ref{kad}) (or its square (\ref{kadop})). Therefore one can
make the whole calculation in the ''e'' coordinatization and in $S^{\prime },
$ the rest frame of the emitter. All results are frame and coordinatization
independent. Now the Doppler effect and the aberration of light are
unseparated phenomena. The results of such calculation agrees with the
experiments \cite{ives} and \cite{mcgow,klein} (made at $\theta =0^{0}$ ($%
180^{0}$)). Also \emph{the ''TT relativity'' calculation predicts the
positive result for the Doppler shift} $\bigtriangleup \lambda $ \emph{(\ref
{dellam}) in the experiments of the type \cite{pobed} and \cite{pobe2}, if
the condition for the aberration angle} $\bigtriangleup \theta $ \emph{(\ref
{aber}) is fulfilled.} This agrees with Bekljamishev's explanation \cite
{bekla} (that is valid only in the ''e'' coordinatization) of the
experiments \cite{pobed} and \cite{pobe2}. The advantage of the ''TT
relativity'' calculation is that it is valid in all permissible
coordinatizations.

The discussion in this paper clearly shows that\emph{\ our invariant
formulation of SR, i.e., the ''TT relativity,'' completely agrees with all
considered experiments in all IFRs and all permissible coordinatizations.}
This is not the case with none of the ''AT relativity'' formulations of SR.
These results are directly contrary to the generally accepted opinion about
the validity of the usual ''AT relativity,'' i.e., of the Einstein
formulation of SR.


\begin{thebibliography}{99}
\bibitem{tom1}  T. Ivezi\'{c}, E-print archives physics/0012048; to be
published in Found. Phys.

\bibitem{tom2}  T. Ivezi\'{c}, E-print archives physics/010191; to be
published in Phys. Essays.

\bibitem{Einst}  A. Einstein, Ann. Physik 17 (1905) 891, tr. by W. Perrett
and G.B. Jeffery, in The principle of relativity, Dover, New York.

\bibitem{ivezic}  T.Ivezi\'{c}, Found. Phys. Lett. 12 (1999) 105.

\bibitem{ive2}  T.Ivezi\'{c}, Found. Phys. Lett. 12 (1999) 507.

\bibitem{rohrl1}  F. Rohrlich, Nuovo Cimento B 45 (1966) 76.

\bibitem{gamba}  A. Gamba, Am. J. Phys. 35 (1967) 83.

\bibitem{Wald}  R.M. Wald, General relativity, The University of Chicago
Press, Chicago, 1984.

\bibitem{Schutz}  B.F. Schutz, A first course in general relativity,
Cambridge University Press, Cambridge, 1985.

\bibitem{Misner}  C.W. Misner, K.S. Thorne and J.A. Wheeler, Gravitation,
Freeman, San Francisco, 1970.

\bibitem{Leub}  C. Leubner, K. Aufinger and P. Krumm, Eur. J. Phys. 13
(1992) 170.

\bibitem{ander}  R. Anderson, I Vetharaniam, G.E. Stedman, Phys. Rep. 295
(1998) 93.

\bibitem{Fahn}  D.E. Fahnline, Am. J. Phys. 50 (1982) 818.

\bibitem{geiger}  K. Geiger, Phys. Rep. 258 (1995) 240.

\bibitem{borch}  V. B\"{o}rchers, J. Meyer, S. Gieseke, G. Martens and C.C.
Noack, Phys. Rev. C 62 (2000) 064903.

\bibitem{ivtruII}  T. Ivezi\'{c}, E-print archives physics/0007031.

\bibitem{caval}  G. Cavalleri and G. Spinelli, Nuovo Cimento B 66 (1970) 11.

\bibitem{gren}  \O . Gr\o n, Am. J. Phys. 49 (1981) 28.

\bibitem{strel}  V. N. Strel'tsov, Found. Phys. 6 (1976) 293; Physics of
Particles and Atomic Nuclei 22 (1991) 1129 (in Russian); Hadronic Journal 17
(1994) 105.

\bibitem{mans}  R. Golestanian, M.R.H. Khajehpour and R. Mansouri, Class.
Quantum Grav. 12 (1995) 273.

\bibitem{Feyn}  R.P. Feynman, R.B. Leightonn and M. Sands, The Feynman
lectures on physics, Vol.1 Addison-Wesley, Reading, 1964 (Sec.15).

\bibitem{Kittel}  C. Kittel, W.D. Knight and M.A. Ruderman, Mechanics,
McGraw-Hill, New York, 1965.

\bibitem{Frisch}  D.H. Frisch and J.H. Smith, Am. J. Phys. 31 (1963) 342.

\bibitem{Easwar}  N. Easwar and D.A. MacIntire, Am. J. Phys. 59 (1991) 589.

\bibitem{rossi}  B. Rossi and D.B. Hall, Phys. Rev. 59 (1941) 223.

\bibitem{ayres}  D.S. Ayres et al., Phys. Rev. D 3 (1971) 1051.

\bibitem{bailey}  J. Bailey et al., Nature 268 (1977) 301; J. Bailey at al.,
Nucl. Phys. B 150 (1979) 1.

\bibitem{newman}  D. Newman, G.W. Ford, A. Rich, and E. Sweetman, Phys. Rev.
Lett. 40 (1978) 1355; F. Combley, F.J.M. Farley, J.H. Field, and E. Picasso,
Phys. Rev. Lett. 42 (1979) 1383; R.D. Sard, Phys. Rev. D. 21 (1980) 549.

\bibitem{huang}  Young-Sea Huang, Helv. Phys.Acta 66\textbf{\ }(1993) 346;
Phys. Essays 9 (1996) 21; Phys. Essays 9 (1996) 340.

\bibitem{field}  J.H. Field, Helv. Phys.Acta 66\textbf{\ }(1993) 875.

\bibitem{nelson}  R.A. Nelson, J. Math. Phys\textit{.} 28 (1987) 2379; J.
Math. Phys\textit{. }35 (1994) 6224\textit{.}

\bibitem{vanzel}  D.A.T. Vanzella and G.E.A. Matsas, H.W. Crater, Am. J.
Phys. 64 (1996) 1075.

\bibitem{ivsc98}  T. Ivezi\'{c}, Preprint SCAN 9802018-CERN.

\bibitem{michels}  A.A. Michelson, E.H. Morley, Am. J. Sci. 34 (1887) 333.

\bibitem{haugan}  M.P. Haugan and C.M. Will, Phys. Today 40\textbf{\ }(1987)
69.

\bibitem{drisc}  R.B. Driscoll, Phys. Essays 10 (1997) 394.

\bibitem{schum}  R.A. Schumacher, Am.J. Phys. 62 (1994) 609.

\bibitem{jaseja}  T.S. Jaseja, A. Javan, J, Murray, and C.H. Townes, Phys.
Rev. A 133 (1964) 1221.

\bibitem{brillet}  A. Brillet and J.L. Hall, Phys. Rev. Lett. 42 (1979) 549.

\bibitem{kenne}  R.J. Kennedy and E.M. Thorndike, Phys. Rev. B. 42 (1932)
400.

\bibitem{hils}  D. Hils and J.L. Hall, Phys. Rev. Lett. 64 (1990) 1697.

\bibitem{ives}  H.E. Ives and G.R. Stilwell, J. Opt. Soc. Am. 28\textbf{\ }%
(1938) 215; 31\textbf{\ }(1941) 369.

\bibitem{mcgow}  R.W. McGowan, D.M. Giltner, S.J. Sternberg S.A. Lee, Phys.
Rev. Lett. 70 (1993) 251.

\bibitem{riis}  E. Riis, U.A. Andersen, N. Bjerre, O.Poulsen, S.A. Lee, and
J.L. Hall, Phys. Rev. Lett. 60 (1988) 81.

\bibitem{klein}  R. Klein et al., Z. Phys. A 342 (1992) 455.

\bibitem{kretz}  M. Kretzschmar, Z. Phys. A - Hadrons and Nuclei 342 (1992)
463).

\bibitem{pobed}  L.A. Pobedonostsev, Y.M. Kramarovsky, P.F. Parshin, B.K.
Seleznev and A.B. Berezin, Journal of Technical Physics 3 (1989) 84 (in
Russian).

\bibitem{pobe2}  L.A. Pobedonostsev, Galilean Electrodynamics 6 (1995) 117.

\bibitem{bekla}  V.O. Bekljamishev, Journal of Technical Physics 69 (1999)
124 (in Russian).
\end{thebibliography}
\end{document}